\pdfoutput=1
\RequirePackage{ifpdf}
\ifpdf 
\documentclass[pdftex]{sigma}
\else
\documentclass{sigma}
\fi

\numberwithin{equation}{section}

\newtheorem{Theorem}{Theorem}[section]
{\theoremstyle{definition}
\newtheorem{Example}[Theorem]{Example}}

\def\C{{\mathbb C}}

\begin{document}

\allowdisplaybreaks

\renewcommand{\thefootnote}{$\star$}

\newcommand{\arXivNumber}{1512.09315}

\renewcommand{\PaperNumber}{038}

\FirstPageHeading

\ShortArticleName{B\^ocher Contractions of Conformally Superintegrable Laplace Equations}

\ArticleName{B\^ocher Contractions of Conformally Superintegrable\\ Laplace Equations\footnote{This paper is a~contribution to the Special Issue on Orthogonal Polynomials, Special Functions and Applications.
The full collection is available at \href{http://www.emis.de/journals/SIGMA/OPSFA2015.html}{http://www.emis.de/journals/SIGMA/OPSFA2015.html}}}

\Author{Ernest G.~KALNINS~$^\dag$, Willard MILLER Jr.~$^\ddag$ and Eyal SUBAG~$^\S$}

\AuthorNameForHeading{E.G.~Kalnins, W.~Miller Jr.\ and E.~Subag}

\Address{$^\dag$~Department of Mathematics, University of Waikato, Hamilton, New Zealand}
\EmailD{\href{mailto:math0236@math.waikato.ac.nz}{math0236@math.waikato.ac.nz}}
\URLaddressD{http://www.math.waikato.ac.nz}

\Address{$^\ddag$~School of Mathematics, University of Minnesota, Minneapolis, Minnesota, 55455, USA}
\EmailD{\href{mailto:miller@ima.umn.edu}{miller@ima.umn.edu}}
\URLaddressD{http://www.ima.umn.edu/~miller/}

\Address{$^\S$~Department of Mathematics, Pennsylvania State University, State College,\\
\hphantom{$^\S$}~Pennsylvania, 16802 USA}
\EmailD{\href{mailto:eus25@psu.edu}{eus25@psu.edu}}

\ArticleDates{Received January 24, 2016, in f\/inal form April 11, 2016; Published online April 19, 2016}

\Abstract{The explicit solvability of quantum superintegrable systems is due to symmetry, but the symmetry is often ``hidden''.
The symmetry generators of 2nd order superintegrable systems in 2 dimensions close under commutation to def\/ine quadratic algebras,
a~generalization of Lie algebras. Distinct systems on constant curvature spaces
are related by geometric limits, induced by generalized In\"on\"u--Wigner Lie algebra contractions of the symmetry
algebras of the underlying spaces.
These have physical/geometric implications, such as the Askey scheme for hypergeometric orthogonal polynomials. However, the limits have no satisfactory Lie algebra contraction interpretations for underlying spaces with~1- or~0-dimensional Lie algebras.
We show that these systems can be best understood by transforming them to Laplace conformally superintegrable systems, with f\/lat space conformal symmetry group ${\rm SO}(4,{\mathbb C})$,
and using ideas introduced in the 1894 thesis of B\^ocher to study separable solutions of the wave equation in terms of roots of
quadratic forms. We show that B\^ocher's prescription for coalescing roots of these forms induces contractions of the
conformal algebra $\mathfrak{so}(4,{\mathbb C})$ to itself and yields a mechanism for classifying all Helmholtz superintegrable systems and their limits.
In the paper [{\it Acta Polytechnica}, to appear, arXiv:1510.09067], we announced our main f\/indings. This paper provides the proofs and more details.}

\Keywords{conformal superintegrability; contractions; Laplace equations}

\Classification{81R05; 81R12; 33C45}

\renewcommand{\thefootnote}{\arabic{footnote}}
\setcounter{footnote}{0}

\section{Introduction}\label{int1}
A quantum (or Helmholtz) superintegrable system is an integrable Hamiltonian system on an $n$-dimensional Riemannian/pseudo-Riemannian manifold
with potential: $H=\Delta_n+V$ that admits $2n-1$
algebraically independent partial dif\/ferential operators $L_j$ commuting with~$H$, the maximum possible.
$ [H,L_j]=0$, $j=1,2,\dots, 2n-1$.
Superintegrability captures the properties of
quantum Hamiltonian systems that allow the Schr\"odinger eigenvalue problem (or Helmholtz equation) $H\Psi=E\Psi$ to be solved exactly, analytically and algebraically~\cite{EVAN, FORDY, MPW2013,TTW2001, SCQS}.
 A~system is of order $K$ if the maximum order of the symmetry
 operators, other than~$H$, is~$K$. For $n=2$, $K=1,2$ all systems are known, e.g.,~\cite{DASK2007, KKM20041,KKM20041II,KKM20041III, KKM20041IV,
 KKM20041V, KKM20041VI}. For $K=1$ the symmetry algebras are just Lie algebras.

 We review quickly the facts for {\it free} 2nd order superintegrable systems (i.e., no potential, $K=2$) in the case $n=2$, $2n-1=3$. The complex spaces with Laplace--Beltrami
 operators admitting at least three 2nd order symmetries were classif\/ied by Koenigs~\cite{Koenigs}. They are:
\begin{itemize}\itemsep=0pt
 \item the two constant curvature spaces (f\/lat space and the complex sphere), six linearly independent 2nd order symmetries and three 1st order symmetries,
 \item the four Darboux spaces (one of which, $D4$, contains a parameter), four 2nd order symmetries and one 1st order symmetry, see Section~\ref{Helmholtzclasses} and \cite{KKMW},
\item 6 families of 4-parameter Koenigs spaces. No 1st order symmetries, see Section~\ref{Helmholtzclasses}.
\end{itemize}
For
2nd order systems with non-constant potential, $K=2$, the generating symmetry operators of
each system close under commutation to determine a
quadratic algebra, and
the irreducible representations of this algebra determine the eigenvalues of $H$ and their multiplicities.
Here we consider only the nondegenerate superintegrable systems. Those with 4-parameter potentials (including the additive constant) (the maximum possible):
\begin{gather*}
V({\bf x})= a_1V_{(1)}({\bf x})+a_2V_{(2)}({\bf x})+a_3V_{(3)}({\bf x})+a_4,
\end{gather*}
where $\{V_{(1)}({\bf x}), V_{(2)}({\bf x}), V_{(3)}({\bf x}), 1\}$ is a linearly independent set.
For these the symmetry algebra generated by $H$, $L_1$, $L_2$ always closes under commutation and gives the following quadratic algebra structure:
Def\/ine the 3rd order commutator $R$ by $R=[L_1,L_2]$. Then
\begin{gather*}[L_j,R]=
A_1^{(j)}L_1^2+A_2^{(j)}L_2^2+A_3^{(j)}H^2+A_4^{(j)}\{L_1,L_2\}+A_5^{(j)}HL_1
+A_6^{(j)}HL_2 \\
\hphantom{[L_j,R]=}{} +A_7^{(j)}L_1+A_8^{(j)}L_2+A_9^{(j)}H+A_{10}^{(j)},\\
 R^2 = b_1 L_1^3 + b_2 L_2^3 + b_3 H^3 + b_4 \big\{L_1^2,L_2\big\} + b_5 \big\{L_1,L_2^2\big\}
 + b_6
L_1 L_2 L_1 + b_7 L_2 L_1 L_2
 \\
\hphantom{R^2 =}{}+ b_8 H\{L_1,L_2\} +b_9 H L_1^2 + b_{10} H L_2^2 + b_{11} H^2
L_1 + b_{12} H^2 L_2 + b_{13} L_1^2 + b_{14} L_2^2 \\
\hphantom{R^2 =}{} + b_{15} \{L_1,L_2\}
+ b_{16} H L_1 + b_{17} H
L_2 + b_{18} H^2 + b_{19} L_1 + b_{20} L_2 + b_{21} H + b_{22},
\end{gather*}
where $\{L_1,L_2\}=L_1L_2+L_2L_1$ and the $A_i^{(j)}$, $b_k$ are constants, dif\/ferent for each algebra.

All 2nd order 2D superintegrable systems with potential and their quadratic algebras are known. There are~33 nondegenerate systems,
on a variety of manifolds, see Section~\ref{Helmholtzclasses} (just manifolds
classif\/ied by Koenigs), where the numbering for constant curvature systems is taken from~\cite{KKMP} (the numbers are not always consecutive because the lists in~\cite{KKMP} also include de\-ge\-ne\-rate systems).
Under the St\"ackel
transform (we discuss this in Section~\ref{2.1}) these systems divide into 8 equivalence classes with representatives on f\/lat space and the 2-sphere, see \cite{Kress2007} and Section~\ref{Stackelequivclasses}.

\subsection{The Helmholtz nondegenerate superintegrable systems} \label{Helmholtzclasses}

{\bf Flat space systems}: $H\Psi=(\partial_x^2+\partial_y^2+V)\Psi=E\Psi$.
\begin{enumerate}\itemsep=0pt
\item $E1$:
$V=\alpha\big(x^2+y^2\big)+\frac{\beta}{x^2}+\frac{\gamma}{y^2}$,

\item $E2$:
$V=\alpha\big(4x^2+y^2\big)+\beta x+\frac{\gamma}{y^2}$,

\item $E3'$: $V=\alpha(x^2+y^2)+\beta x+\gamma y$,

 \item $E7$:
$V=\frac{\alpha(x+iy)}{\sqrt{(x+iy)^2-b}}+\frac{\beta (x-iy)}{\sqrt{(x+iy)^2-b}\left(x+iy+\sqrt{(x+iy)^2-b}\right)^2}+\gamma \big(x^2+y^2\big)$,

\item $E8$: $V=\frac{\alpha (x-iy)}{(x+iy)^3}+\frac{\beta}{(x+iy)^2}+\gamma\big(x^2+y^2\big)$,

\item $E9$:
$V=\frac{\alpha}{\sqrt{x+iy}}+\beta y+\frac{\gamma (x+2iy)}{\sqrt{x+iy}}$,

\item $E10$:
$V=\alpha(x-iy)+\beta \big(x+iy-\frac32(x-iy)^2\big)+\gamma\big(x^2+y^2-\frac12(x-iy)^3\big)$,

\item $E11$:
$V=\alpha(x-iy)+\frac{\beta (x-iy)}{\sqrt{x+iy}}+\frac{\gamma}{\sqrt{x+iy}}$,

\item $E15$: \looseness=-1
$V=f(x-iy)$, where $f$ is arbitrary (the exceptional case, characterized by the~fact that the symmetry generators
 are functionally linearly dependent  \cite{KKM20041,KKM20041II,KKM20041III, KKM20041IV,
 KKM20041V, KKM20041VI, KKMP}),

 \item $E16$: $V=\frac{1}{\sqrt{x^2+y^2}}\left(\alpha+\frac{\beta}{y+\sqrt{x^2+y^2}}+\frac{\gamma}{y-\sqrt{x^2+y^2}}\right)$,

\item $E17$:
$V=\frac{\alpha}{\sqrt{x^2+y^2}}+\frac{\beta}{(x+iy)^2}+\frac{\gamma}{(x+iy)\sqrt{x^2+y^2}}$,

\item $E19$:
$V=\frac{\alpha(x+iy)}{\sqrt{(x+iy)^2-4}}+\frac{\beta}{\sqrt{(x-iy)(x+iy+2)}}+\frac{\gamma}{\sqrt{(x-iy)(x+iy-2)}}$,

\item $E20$:
$V=\frac{1}{\sqrt{x^2+y^2}}\left(\alpha+\beta \sqrt{x+\sqrt{x^2+y^2}}+\gamma \sqrt{x-\sqrt{x^2+y^2}}\right)$.
\end{enumerate}

{\bf Systems on the complex 2-sphere}: $H\Psi=(J_{23}^2+J_{13}^2+J_{12}^2+V)\Psi=E\Psi$.
Here, $J_{k\ell}=s_k\partial_{s_\ell}-s_\ell\partial_{s_k}$ and $s_1^2+s_2^2+s_3^2=1$.
\begin{enumerate}\itemsep=0pt

\item $S1$:
$V=\frac{\alpha}{(s_1+is_2)^2}+\frac{\beta s_3}{(s_1+is_2)^2}+\frac{\gamma(1-4s_3^2)}{(s_1+is_2)^4}$,

\item $S2$:
$V=\frac{\alpha}{s_3^2}+\frac{\beta}{(s_1+is_2)^2}+\frac{\gamma(s_1-is_2)}{(s_1+is_2)^3}$,

\item $S4$:
$V=\frac{\alpha}{(s_1+is_2)^2}+\frac{\beta s_3}{\sqrt{s_1^2+s_2^2}}+\frac{\gamma}{(s_1+is_2)\sqrt{s_1^2+s_2^2}}$,

\item $S7$:
$V=\frac{\alpha s_3}{\sqrt{s_1^2+s_2^2}}+\frac{\beta s_1}{s_2^2\sqrt{s_1^2+s_2^2}}+\frac{\gamma}{s_2^2}$,

\item $S8$:
$V=\frac{\alpha s_2}{\sqrt{s_1^2+s_3^2}}+\frac{\beta (s_2+is_1+s_3)}{\sqrt{(s_2+is_1)(s_3+is_1)}}+\frac{\gamma(s_2+is_1-s_3)}{\sqrt{(s_2+is_1)(s_3-is_1)}}$,

\item $S9$:
$V=\frac{\alpha}{s_1^2}+\frac{\beta}{s_2^2}+\frac{\gamma}{s_3^2}$.
\end{enumerate}

{\bf Darboux 1 systems}: $H\Psi=\left(\frac{1}{4x}(\partial_x^2+\partial_y^2)+V\right)\Psi=E\Psi$.
\begin{enumerate}\itemsep=0pt
\item ${D1A}$:
$ V=\frac{b_1(2x-2b+iy)}{x\sqrt{x-b+iy}}+\frac{b_2}{x\sqrt{x-b+iy}}+\frac{b_3}{x}+b_4$,
\item ${D1B}$: $V=\frac{b_1(4x^2+y^2)}{x}+\frac{b_2}{x}+\frac{b_3}{xy^2}+b_4$,

 \item ${D1C}$: $V=\frac{b_1(x^2+y^2)}{x}+\frac{b_2}{x}+\frac{b_3y}{x}+b_4$.
\end{enumerate}

{\bf Darboux 2 systems}: $H\Psi=\left(\frac{x^2}{x^2+1}(\partial_x^2+\partial_y^2)+V\right)\Psi=E\Psi$.
\begin{enumerate}\itemsep=0pt
\item ${D2A}$: $V=\frac{x^2}{x^2+1}\left(b_1(x^2+4y^2)+\frac{b_2}{x^2}+b_3y\right)+b_4$,
\item ${D2B}$: $V=\frac{x^2}{x^2+1}\left(b_1(x^2+y^2)+\frac{b_2}{x^2}+\frac{b_3}{y^2}\right)+b_4$,
 \item ${D2C}$:
$V=\frac{x^2}{\sqrt{x^2+y^2}(x^2+1)}\left(b_1+\frac{b_2}{y+\sqrt{x^2+y^2}}+\frac{b_3}{y-\sqrt{x^2+y^2}}\right)+b_4$.
\end{enumerate}

{\bf Darboux 3 systems}: $H\Psi=\left(\frac12\frac{e^{2x}}{e^x+1}(\partial_x^2+\partial_y^2)+V\right)\Psi=E\Psi$.
\begin{enumerate}\itemsep=0pt
 \item ${D3A}$: $V=\frac{b_1}{1+e^x}+\frac{b_2e^x}{\sqrt{1+2e^{x+iy}}(1+e^x)}+\frac{b_3e^{x+iy}}{\sqrt{1+2e^{x+iy}}(1+e^x)}+b_4$,
\item ${D3B}$: $V=\frac{e^x}{e^x+1}\left(b_1+e^{-\frac{x}{2}}(b_2\cos\frac{y}{2}+b_3\sin\frac{y}{2})\right)+b_4$,
 \item ${D3C}$: $V(= \frac{e^x}{e^x+1}\left(b_1+e^x(\frac{b_2}{\cos^2\frac{y}{2}}+\frac{b_3}{\sin^2\frac{y}{2}})\right)+b_4$,
 \item ${D3D}$: $V=\frac{e^{2x}}{1+e^x}\big(b_1 e^{-iy}+b_2 e^{-2iy}\big)+\frac{b_3}{1+e^x}+b_4$.
\end{enumerate}

{\bf Darboux 4 systems}: $H\Psi=\left(-\frac{\sin^2 2x}{2\cos 2x+b}(\partial_x^2+\partial_y^2)+V\right)\Psi=E\Psi$.
\begin{enumerate}\itemsep=0pt
\item ${D4(b)A}$: $V=\frac{\sin^2 2 x}{2 \cos 2 x+b}\left(\frac{b_1}{\sinh^2 y}+\frac{b_2}{\sinh^2 2 y}\right)
+\frac{b_3}{2 \cos 2 x+b}+b_4$,
\item $D4(b)B$: $V=\frac{\sin^2 2x}{2 \cos 2x +b}\left(\frac{b_1}{\sin^2 2x}+b_2e^{4y}+b_3e^{2y}\right)+b_4$,
\item ${D4(b)C}$: $V=\frac{e^{2y}}{\frac{b+2}{\sin^2 x}+\frac{b-2}{\cos^2 x}}\left(\frac{b_1}{Z+(1-e^{2y})\sqrt{Z}}+\frac{b_2}{Z+(1+e^{2y})\sqrt{Z}}
+\frac{b_3e^{-2y}}{\cos^2 x}\right) +b_4$,\\
$Z=(1-e^{2y})^2+4e^{2y}\cos^2x$.
\end{enumerate}

{\bf Generic Koenigs spaces}:
\begin{enumerate}\itemsep=0pt \item ${K[1,1,1,1]}$:
$H\Psi=\frac{1}{V(b_1,b_2,b_3,b_4)}\big(\partial_x^2+\partial_y^2 +V(a_1,a_2,a_3,a_4)\big)\Psi=E\Psi$,\\
$V(a_1,a_2,a_3,a_4)=\frac{a_1}{x^2}+\frac{a_2}{y^2}+\frac{4a_3}{(x^2+y^2-1)^2}-\frac{4a_4}{(x^2+y^2+1)^2}$,
 \item ${K[2,1,1]}$:
$H\Psi=\frac{1}{V(b_1,b_2,b_3,b_4)}\big(\partial_x^2+\partial_y^2 +V(a_1,a_2,a_3,a_4)\big)\Psi=E\Psi$,\\
 $V(a_1,a_2,a_3,a_4)=\frac{a_1}{x^2}+\frac{a_2}{y^2}-a_3\big(x^2+y^2\big)+a_4$,
 \item ${K[2,2]}$:
$H\Psi=\frac{1}{V(b_1,b_2,b_3,b_4)}\big(\partial_x^2+\partial_y^2 +V(a_1,a_2,a_3,a_4)\big)\Psi=E\Psi$,\\
$V(a_1,a_2,a_3,a_4)= \frac{a_1}{(x+iy)^2}+\frac{a_2(x-iy)}{(x+iy)^3}
+a_3-a_4\big(x^2+y^2\big)$,
\item ${K[3,1]}$:
$H\Psi=\frac{1}{V(b_1,b_2,b_3,b_4)}\big(\partial_x^2+\partial_y^2 +V(a_1,a_2,a_3,a_4)\big)\Psi=E\Psi$,\\
$V(a_1,a_2,a_3,a_4)= a_1-a_2x
+a_3\big(4x^2+{y}^2\big)+\frac{a_4}{{y}^2}$,
\item ${K[4]}$:
$H\Psi=\frac{1}{V(b_1,b_2,b_3,b_4)}\big(\partial_x^2+\partial_y^2 +V(a_1,a_2,a_3,a_4)\big)\Psi=E\Psi$,\\
$V(a_1,a_2,a_3,a_4)= a_1-a_2(x+iy)
+a_3\big(3(x+iy)^2+2(x-iy)\big)
-a_4\big(4(x^2+y^2)+2(x+iy)^3\big)$,
\item ${K[0]}$:
$H\Psi=\frac{1}{V(b_1,b_2,b_3,b_4)}\big(\partial_x^2+\partial_y^2 +V(a_1,a_2,a_3,a_4)\big)\Psi=E\Psi$,\\
$V(a_1,a_2,a_3,a_4)=a_1-(a_2x+a_3y)+a_4\big(x^2+y^2\big)$.
\end{enumerate}

\subsection{Lie algebras and quadratic algebras}\label{1.2}

Important for 2nd order superintegrable systems are the Lie algebras $\mathfrak{e}(2,\C)$ and $\mathfrak{o}(3,\C)$.
These are the ($K=1$) symmetry Lie algebras of free (zero potential) systems on constant curvature spaces.
 Every 2nd order symmetry operator on a constant curvature space takes the form
$ L= K + W({\bf x})$,
where $K$ is a 2nd order element in the enveloping algebra of $\mathfrak{o}(3,\C)$ or $\mathfrak{e}(2,\C)$.
An important example is $S_9$:
 \begin{gather*}
 H=J_1^2+J_2^2+J_3^2+\frac{a_1}{s_1^2}+\frac{a_2}{s_2^2}+\frac{a_3}{s_3^2},
 \end{gather*}
where $J_3=s_1\partial_{s_2}-s_2\partial_{s_1}$ and $J_2$, $J_3$
are obtained by cyclic permutations of indices.
Basis symmetries are
\begin{gather*}
L_1=J_1^2+\frac{a_3 s_2^2}{s_3^2}+\frac{a_2 s_3^2}{s_2^2},\qquad
 L_2=J_2^2+\frac{a_1 s_3^2}{s_1^2}+\frac{a_3 s_1^2}{s_3^2},\qquad
 L_3=J_3^2+\frac{a_2 s_1^2}{s_2^2}+\frac{a_1 s_2^2}{s_1^2}.
 \end{gather*}

 \begin{Theorem}\label{theorem1} There is a bijection between quadratic algebras generated by $2$nd order elements in
 the enveloping algebra of $\mathfrak{o}(3,\C)$, called free, and $2$nd order nondegenerate superintegrable systems on the
 complex $2$-sphere. Similarly, there is a bijection between quadratic algebras generated by $2$nd order elements in
 the enveloping algebra of $\mathfrak{e}(2,\C)$ and $2$nd order nondegenerate superintegrable systems on the $2D$ complex flat space.
 \end{Theorem}

The proof of this theorem is constructive~\cite{KM2014}. Given a free quadratic algebra $\tilde Q$ one can compute
the potential~$V$ and
the symmetries of the quadratic algebra $Q$ of the nondegenerate superintegrable system.
These systems are closely related to the special functions of mathematical physics and their properties.
The special functions arise in two distinct ways: 1)
 As separable eigenfunctions of the quantum Hamiltonian. Second order superintegrable systems are multiseparable,
 (with one exception) \cite{KKM20041,KKM20041II,KKM20041III, KKM20041IV,
 KKM20041V, KKM20041VI}.
 2) As interbasis expansion coef\/f\/icients relating distinct separable coordinate eigenbases~\cite{KMP2007a,KMP2008,LM2014,P2011}.
 Most of the classical special functions in the Digital Library of Mathematical Functions, as well as
 Wilson polynomials, appear in these ways~\cite{DLMF}.

In \cite{KM2014} it has been shown that
 all the 2nd order superintegrable systems are obtained by taking coordinate limits of the generic system~$S_9$~\cite{KKMP}, or are obtained from these limits by a~St\"ackel
transform (an invertible structure preserving mapping of superintegrable systems~\cite{KKM20041,KKM20041II,KKM20041III, KKM20041IV,
 KKM20041V, KKM20041VI}). Analogously all
quadratic symmetry algebras of these
systems are limits of that of~$S_9$. These coordinate limits induce limit relations between the special functions
associated as eigenfunctions of the superintegrable systems.
 The limits also induce contractions of the associated quadratic algebras, and via the models of the irreducible representations of
 these algebras, limit relations between the associated
 special functions. The Askey scheme for ortho\-go\-nal functions of hypergeometric type is an example of this~\cite{KMP2014}.
 For constant curvature systems the required limits are all induced by In\"on\"u--Wigner-type Lie algebra contractions of~$\mathfrak{o}(3,\C)$ and~$\mathfrak{e}(2,\C)$~\cite{Wigner,NP,WW}. In\"on\"u--Wigner-type Lie algebra contractions have long been applied to
 relate separable coordinate systems
 and their associated special functions, see, e.g.,~\cite{Pog01, Pog96} for some more recent examples, but the application to quadratic algebras
 is due to the authors and their collaborators.

Recall the def\/inition of (natural) {\it Lie algebra contractions}:
Let $(A; [\, ;\, ]_A)$, $(B; [ \,; \,]_B)$ be two complex Lie algebras. We say that
$B$ is a~{\it contraction} of $A$ if for every $\epsilon\in (0; 1]$ there exists a~linear invertible
map $t_\epsilon \colon B\to A$ such that for every $X, Y\in B$,
$ \lim\limits_{\epsilon\to 0}t_\epsilon^{-1}[t_\epsilon X,t_\epsilon Y]_A
= [X, Y ]_B$.
Thus, as $\epsilon\to 0$ the 1-parameter family of basis transformations can become nonsingular but
the structure constants of the Lie algebra go to a f\/inite limit, necessarily that of another Lie algebra.

The contractions of the symmetry algebras of constant curvature spaces have long since been classif\/ied~\cite{KM2014}. There are 6 nontrivial contractions of
$\mathfrak{e}(2,\C)$ and 4 of $\mathfrak{o}(3,\C)$. They are each induced by coordinate limits.

{\it Contractions of quadratic algebras:}
Just as for Lie algebras we can def\/ine a contraction of a~quadratic algebra in terms of 1-parameter families of basis changes
in the algebra:
As $\epsilon\to 0$ the 1-parameter family of basis transformations becomes singular but the structure constants go to a~f\/inite limit~\cite{KM2014}.

Let $H=H^{(0)}+V$, $S_1=S^{(0)}_1+W_1$, $S_2=S^{(0)}_2+W_2$ be a superintegrable system on a constant curvature space with
quadratic algebra $Q$ and free quadratic
algebra $\tilde Q$ of $H^{(0)}$, $S^{(0)}_1$, $S^{(0)}_2$.
Motivating idea: {\it Lie algebra contractions induce quadratic algebra contractions.}
For constant curvature spaces we have
\begin{Theorem}[\cite{KM2014}] Every Lie algebra contraction of $A=\mathfrak{e}(2,\C)$ or $A=\mathfrak{o}(3,\C)$ induces a~contraction of a free
$($zero potential$)$ quadratic
 algebra~$\tilde Q$ based on~$A$,
which in turn induces a~contraction of the quadratic algebra $Q$ with potential. This is true for both classical and quantum algebras.
\end{Theorem}

\looseness=-1
Similarly the coordinate limit associated with each contraction takes $H$ to a new superintegrable system with
the contracted quadratic algebra.
This relationship between coordinate limits, Lie algebra contractions and quadratic algebra contractions
for superintegrable systems on constant curvature spaces breaks down for Darboux and Koenigs spaces.
For Darboux spaces the Lie symmetry algebra is only 1-dimensional, and there is no Lie symmetry algebra at all for
Koenigs spaces. Furthermore, there is the issue of f\/inding a more systematic way of classifying the
44 distinct Helmholtz superintegrable eigenvalue systems on dif\/ferent manifolds, and their relations.
These issues can be clarif\/ied by considering the Helmholtz systems as Laplace equations (with potential) on f\/lat space.
This point of view was introduced in the paper~\cite{Laplace2011} and applied in~\cite{CapelKress} to solve important classif\/ication
problems in the case $n=3$.
As announced in~\cite{KMS2016}, the proper object to study is the conformal symmetry algebra $\mathfrak{so}(4,C)$ of the f\/lat space Laplacian
and its contractions. The basic idea is that families of (St\"ackel-equivalent) Helmholtz
superintegrable systems
on a variety of manifolds correspond to a single conformally superintegrable Laplace equation on f\/lat space. We exploit this
relation in the case $n=2$,
but it generalizes easily to all dimensions $n\ge 2$. The conformal symmetry algebra for Laplace equations with
constant potential on f\/lat space is the
conformal algebra $\mathfrak{so}(n+2,\C)$. We review these concepts in Section~\ref{2}.

In his 1894 thesis \cite{Bocher} B\^ocher introduced a limit procedure based on the roots of quadratic forms to f\/ind families of
$R$-separable solutions of the ordinary (zero potential) f\/lat space Laplace equation in $n$ dimensions. An important feature of his work was the
introduction of special projective coordinates in which the action of the conformal group $\mathfrak{so}(n+2,\C)$ on solutions of the Laplace equation can be linearized. For $n=2$ these are tetraspherical coordinates. In Section~\ref{3} we describe in detail the Laplace equation
mechanism and how it can be applied to systematize the classif\/ication of Helmholtz
superintegrable systems and their relations via limits. We show that B\^ocher's limit procedure can be interpreted as constructing
generalized In\"on\"u--Wigner Lie algebra contractions of $\mathfrak{so}(4,\C)$ to itself. We call these B\^ocher contractions and show that they induce contractions of the conformal quadratic algebras associated with Laplace superintegrable systems. All of the
limits of the Helmholtz systems classif\/ied before for $n=2$~\cite{KM2014} are induced by the larger class of B\^ocher contractions.

\section{2D conformal superintegrability of the 2nd order} \label{2}

Systems of Laplace type are of the form
 \begin{gather}
 \label{Laplace} H\Psi\equiv \Delta_n\Psi+V\Psi=0.
 \end{gather}
 Here $\Delta_n$ is the Laplace--Beltrami operator on a real or complex conformally f\/lat~$nD$ Riemannian or pseudo-Riemannian manifold. We assume that all functions occurring in this paper are
 locally analytic, real or complex.)
A conformal symmetry of this equation is a partial dif\/ferential operator $S$ in the variables ${\bf x}=(x_1,\dots,x_n)$ such
that $[ S, H]\equiv SH-HS=R_{S} H$ for some dif\/ferential operator~$R_{S}$.
A conformal symmetry maps any solution~$\Psi$ of~(\ref{Laplace}) to another solution. Two conformal symmetries ${S}$, ${S}'$ are
identif\/ied if $S=S'+RH$ for some dif\/ferential operator $R$, since they agree on the solution space of~(\ref{Laplace}). (For short we will say
that \mbox{$S=S'$}, $\operatorname{mod} (H)$
and that $S$ is a symmetry if $[S,H]=0$, $\operatorname{mod}(H)$.)
The system is {\it conformally superintegrable} for $n>2$ if there are $2n-1$ functionally independent conformal symmetries,
${S}_1,\dots,{S}_{2n-1}$ with ${S}_1={H}$. It is {\it second order conformally superintegrable} if each
symmetry $S_i$ can be chosen to be a dif\/ferential operator of at most second order.

For $n=2$ the def\/inition must be restricted, since for a potential $V=0$ there will be an inf\/inite-dimensional space of conformal
symmetries when $n=2$; every analytic function induces such symmetries.
We assume $V\ne 0$, possibly a constant.

Every $2D$ Riemannian manifold is conformally f\/lat, so we can always f\/ind a Cartesian-like coordinate system with coordinates ${\bf x}\equiv (x,y)\equiv (x_1,x_2)$
such that the Laplace equation takes the form
\begin{gather}
\label{Laplace4} {\tilde H}=\frac{1}{\lambda(x,y)}\big(\partial_x^2+\partial_y^2\big)+{\tilde V}({\bf x})=0.
\end{gather}
However, this equation is equivalent to the f\/lat space equation
\begin{gather}
\label{Laplace5}{H}\equiv \partial_x^2+\partial_y^2+ V({\bf x})=0,\qquad V({\bf x})=\lambda({\bf x}){\tilde V}({\bf x}).
\end{gather}
In particular, the conformal symmetries of (\ref{Laplace4}) are identical with the conformal symmetries of~(\ref{Laplace5}).
Indeed, denoting by $\Lambda$ the operator of multiplication by the function $\lambda(x,y)$ and using the operator identity $[A,BC]=B[A,C]+[A,B]C$ we have
\begin{gather*}
[S,H]=[S,\Lambda{\tilde H}] =\Lambda[S,{\tilde H}]+[S,\Lambda]{\tilde H}=
\Lambda R{\tilde H}+[S,\Lambda]{\tilde H}=\big(\Lambda R \Lambda^{-1}+[S,\Lambda]\Lambda^{-1}\big)H.
\end{gather*}
Thus without loss of generality we can assume the manifold is f\/lat space with $\lambda\equiv 1$.

Since the Hamiltonians are formally self-adjoint, without loss of generality we can always assume that a 2nd order
conformal symmetry $S$ is formally self-adjoint:
 \begin{gather*}
{S}=\frac{1}{\lambda}\sum ^2_{k,j=1}\partial_k\cdot \big(\lambda a^{kj}({\bf x})\big)\partial_j +W({\bf x})\equiv
S_0+W,\qquad a^{jk}=a^{kj}. 
\end{gather*}
Equating coef\/f\/icients of the partial derivatives on both sides of
\begin{gather*}
[S,H]=\big(R^{(1)}({\bf x})\partial_x+R^{(2)}({\bf x})\partial_y\big)H,
\end{gather*}
we can derive the conditions
\begin{gather*}
a_i^{ii}=2a_j^{ij}+a_i^{jj},\qquad i\ne j,\qquad W_j=\sum_{s=1}^2a^{sj}V_s+a_j^{jj}V,\qquad k,j=1,2.
\end{gather*}
Here a subscript $j$ on $a^{\ell m}$, $V$ or $W$ denotes dif\/ferentiation with respect~to $x_j$.
The requirement that $\partial_{x} W_2=\partial_{y}W_1$ leads
to the
second order (conformal) Bertrand--Darboux partial dif\/ferential equation for the potential
\begin{gather*}
a^{12}(V_{11}-V_{22})+\big(a^{22}-a^{11}\big)V_{12}+\big(a^{12}_1+a^{22}_2-a^{11}_2\big)V_1+\big(a^{22}_1-a^{11}_1-a^{12}_2\big)V_2+2a^{12}_{12}V=0.
\end{gather*}
The following results are easy modif\/ications of results for 3D conformal superintegrable systems proved in~\cite{Laplace2011}.
For a conformally superintegrable system there are 3 functionally independent symmetries, each leading to a~Bertrand--Darboux equation. The result is that the potential
function $V$ must satisfy a canonical system of equations of the form
\begin{gather}
 V_{22} = V_{11} + A^{22}({\bf x}) V_1+B^{22}({\bf x}) V_2+C^{22}({\bf x}) V,\nonumber\\
 V_{12} = A^{12}({\bf x}) V_1+B^{12}({\bf x}) V_2+C^{12}({\bf x}) V.\label{nondegpot1}
\end{gather}
If the integrability conditions for this system~(\ref{nondegpot1}) are satisf\/ied identically, the vector space of solutions
$V$ is four-dimensional and we say
that the potential is {\it nondegenerate}. Otherwise the potential is {\it degenerate} and the potential involves $\le 3$ parameters.
 In this paper we consider only systems with nondegenerate potentials.
Since we can always add the trivial conformal symmet\-ry~$\rho({\bf x})H$ to~$S$ we could assume that, say $a_{11}=0$.

In general the space of 2nd order conformal symmetries could be inf\/inite-dimensional. However, the requirement that
$H$ have a multiparameter potential
reduces the possible symmetries to a f\/inite-dimensional space. Indeed the conformal Bertrand--Darboux conditions
for a 2nd order symmetry yields the requirement $\partial_{xy}(a^{11}-a^{22})=0$.
The result is that the pure derivative terms $S_0$ belong to the space
spanned by symmetrized products of the conformal Killing vectors
\begin{gather}
P_1=\partial_x,\qquad P_2=\partial_y,\qquad J=x\partial_y-y\partial_x,\qquad
D=x\partial_x+y\partial_y,\nonumber\\
K_1=\big(x^2-y^2\big)\partial_x +2xy\partial_y, \qquad K_2=\big(y^2-x^2\big)\partial_y+2xy\partial_x. \label{conformalKV}
\end{gather}
and terms $g({\bf x})\big(\partial_x^2+\partial_y^2\big)$, where $g$ is an
arbitrary function. For a given multiparameter potential only a subspace of these conformal tensors occurs.
Note that on the hypersurface ${\cal H}=0$ in phase space all symmetries $g({\bf x}){\cal H}$ vanish, so any
two symmetries dif\/fering by $g({\bf x}){\cal H}$ can be identif\/ied.

\subsection{The conformal St\"ackel transform} \label{2.1}
We review quickly the concept of the St\"ackel transform \cite{CCM} and extend it to conformally superintegrable systems.
Suppose we have a second order {\it conformal} superintegrable system
\begin{gather}
\label{confl} {H}=\frac{1}{\lambda(x,y)}(\partial_{xx}+\partial_{yy})+V(x,y)=0,\qquad {H}={H}_0+V
\end{gather}
with $V$ the general solution for this system, and suppose $U(x,y)$ is a particular potential solution, nonzero in an open set. The {\it St\"ackel transform} induced by~$U$ is the system
\begin{gather}\label{helms} {\tilde {H}}=E,\quad
{\tilde {H}}=\frac{1}{{\tilde \lambda}}(\partial_{xx}+\partial_{yy})+{\tilde V}, \qquad {\rm where} \qquad
{\tilde \lambda}=\lambda U, \qquad {\tilde V}=\frac{V}{U}.
\end{gather}
In \cite{Laplace2011,KMS2016} we proved
\begin{Theorem}\label{stackelt}
The transformed $($Helmholtz$)$ system $\tilde H$
is truly superintegrable.
\end{Theorem}

Note that if $H\Psi=0$ then ${\tilde S}\Psi =S\Psi$ and $H(S\Psi)=0$ so $S$ and $\tilde S$ agree on the null space of~$H$ and
they preserve this null space.
This result shows that any second order conformal Laplace superintegrable system admitting a nonconstant potential~$U$ can be
St\"ackel transformed to a~Helmholtz
superintegrable system. This operation is invertible, but the inverse is not a St\"ackel transform.
By choosing all possible special potentials $U$ associated with the f\/ixed Laplace system~(\ref{confl}) we generate the equivalence class of all Helmholtz
superintegrable systems~(\ref{helms}) obtainable through this process. As is easy to check, any two Helmholtz superintegrable systems lie in the same equivalence class
if and only if they are St\"ackel equivalent in the standard sense, see~\cite[Theorem~4]{KMS2016}. All Helmholtz superintegrable systems are related to conformal Laplace systems in this way,
so the study of all Helmholtz superintegrability on conformally f\/lat manifolds can be reduced to the study of all conformal Laplace
superintegrable systems on f\/lat space.

In \cite{KKM20041,KKM20041II,KKM20041III, KKM20041IV, KKM20041V, KKM20041VI} it is demonstrated that for the 3-parameter Helmholtz system~$H'$ and the St\"ackel transform ${\tilde H}'$,
\begin{gather*}
H'=H_0+V'=H_0+U^{(1)}\alpha_1+U^{(2)}\alpha_2+U^{(3)}\alpha_3,\\
{\tilde H}'=\frac{1}{U^{(1)}}H_0
 +\frac{-U^{(1)}E+U^{(2)}\alpha_2+U^{(3)}\alpha_3}{U^{(1)}},
\end{gather*}
if $H'\Psi=E\Psi$ then ${\tilde H}'\Psi=-\alpha_1\Psi$. The ef\/fect of the St\"ackel transform is
to replace~$\alpha_1$ by~$-E$ and~$E$
by~$-\alpha_1$. Further, a~2nd order symmetry~$S$ of~$H'$ transforms to the 2nd order symmetry~$\tilde S$ of~${\tilde H}'$
such that~$S$ and~$\tilde S$ agree on eigenspaces of~$H'$.

We know that the symmetry operators of all 2nd order nondegenerate superintegrable systems in 2D generate a quadratic algebra
of the form
\begin{gather}
 [R,S_1]=f^{(1)}(S_1,S_2,\alpha_1,\alpha_2,\alpha_3,H'),\qquad [R,S_2]=f^{(2)}(S_1,S_2,\alpha_1,\alpha_2,\alpha_3,H'),\nonumber\\
 \label{quadratic1} R^2=f^{(3)}(S_1,S_2,\alpha_1,\alpha_2,\alpha_3,H'),\qquad R\equiv [S_1,S_2],
 \end{gather}
where $\{S_1,S_2,H\}$ is a basis for the 2nd order symmetries and $\alpha_1$, $\alpha_2$, $\alpha_3$ are the parameters
for the potential \cite{KKM20041,KKM20041II,KKM20041III, KKM20041IV,
 KKM20041V, KKM20041VI,MPW2013}. We see that the ef\/fect of a St\"ackel transform generated by the potential function~$U^{(1)}$ is to determine a~new superintegrable
system with structure
\begin{gather} [{\tilde R},{\tilde S}_1]=f^{(1)}\big({\tilde S}_1,{\tilde S}_2,-{\tilde H}',\alpha_2,\alpha_3,-\alpha_1\big),\qquad
[{\tilde R},{\tilde S}_2]=f^{(2)}\big({\tilde S}_1,{\tilde S}_2,-{\tilde H}',\alpha_2,\alpha_3,-\alpha_1\big),\nonumber\\
 \label{quadratic2}
 {\tilde R}^2=f^{(3)}\big({\tilde S}_1,{\tilde S}_2,-{\tilde H}',\alpha_2,\alpha_3,-\alpha_1\big),\qquad {\tilde R}\equiv [{\tilde S}_1,{\tilde S}_2].
 \end{gather}
Of course, the switch of $\alpha_1$ and $H'$ is only for illustration; there is a St\"ackel transform that replaces any
$\alpha_j$ by $-H'$ and $H'$ by $-\alpha_j$.

Formulas (\ref{quadratic1}) and (\ref{quadratic2}) are just instances of the quadratic algebras of the superintegrable systems belonging to the
equivalence class of a single nondegenerate conformally superintegrable Hamiltonian
$\hat{H}=\partial_{xx}+\partial_{yy}+\sum\limits_{j=1}^4\alpha_j V^{(j)}(x,y)$.
Let $\hat{S}_1$, $\hat{S}_2$, $\hat{H}$ be a basis of 2nd order conformal symmetries of $\hat H$. From the above discussion we can conclude the following.

\begin{Theorem} The symmetries of the $2D$ nondegenerate conformal superintegrable Hamiltonian~$\hat H$ generate a quadratic algebra
 \begin{gather}
 [{\hat R},{\hat S}_1]=f^{(1)}\big({\hat S}_1,\hat{S}_2,\alpha_1,\alpha_2,\alpha_3,\alpha_4\big),\qquad [{\hat R},{\hat S}_2]=f^{(2)}
 \big({\hat S}_1,{\hat S}_2,\alpha_1,\alpha_2,\alpha_3,\alpha_4\big),\nonumber\\
{\hat R}^2=f^{(3)}\big({\hat S}_1,\hat{S}_2,\alpha_1,\alpha_2,\alpha_3,\alpha_4\big),\label{confquadalg}
\end{gather}
where $\hat{R}=[{\hat S}_1,\hat{S}_2]$ and all identities hold $\operatorname{mod}({\hat H})$. A conformal St\"ackel transform generated by the potential
$V^{(j)}(x,y)$ yields a nondegenerate Helmholtz superintegrable Hamiltonian $\tilde H$ with quadratic algebra relations identical to~\eqref{confquadalg},
except that we make the replacements ${\hat S}_\ell\to {\tilde S}_\ell$ for $\ell=1,2$ and $\alpha_j\to -{\tilde H}$. These modified relations~\eqref{confquadalg} are now true identities, not $\operatorname{mod} ({\hat H})$.
\end{Theorem}

\section{Tetraspherical coordinates} \label{Tc}
The tetraspherical coordinates $(x_1,\dots,x_4)$ satisfy $x_1^2+x_2^2+x_3^2+x_4^2=0$ (the null cone) and $\sum\limits_{k=1}^4x_k\partial_{x_k}=0$.
They are projective coordinates on the null cone and have 3 degrees of freedom. Their principal advantage over f\/lat space Cartesian coordinates is
that the action of the conformal algebra~(\ref{conformalKV}) and of the conformal group $\sim {\rm SO}(4,\C)$ is linearized in tetraspherical coordinates.

 {\bf Relation to Cartesian coordinates $(x,y)$ and coordinates on the 2-sphere $(s_1,s_2,s_3)$}:
\begin{gather*} x_1=2XT,\qquad x_2=2YT,\qquad x_3=X^2+Y^2-T^2,\qquad x_4=i\big(X^2+Y^2+T^2\big),\\
 x=\frac{X}{T}=-\frac{x_1}{x_3+ix_4},\qquad y=\frac{Y}{T}=-\frac{x_2}{x_3+ix_4},\qquad
 x=\frac{s_1}{1+s_3},\qquad y=\frac{s_2}{1+s_3},\\
 s_1=\frac{2x}{x^2+y^2+1},\qquad s_2=\frac{2y}{x^2+y^2+1},\qquad s_3=\frac{1-x^2-y^2}{x^2+y^2+1},\\
 H=\partial_{xx}+\partial_{yy}+{\tilde V}=(x_3+ix_4)^2\left(\sum_{k=1}^4\partial_{x_k}^2+V\right)
=(1+s_3)^2\left(\sum_{j=1}^3p_{s_j}^2+V\right),\\
 {\tilde V}=(x_3+ix_4)^2V, \qquad
 (1+s_3)=-i\frac{(x_3+ix_4)}{x_4},\\
s_1=\frac{ix_1}{x_4},\qquad s_2=\frac{ix_2}{x_4},\qquad s_3=\frac{-ix_3}{x_4}.
\end{gather*}

\noindent {\bf Relation to f\/lat space and 2-sphere 1st order conformal constants of the motion}:
We def\/ine
\begin{gather*}
L_{jk}=x_j\partial_{x_k}-x_k \partial_{x_j},\qquad 1\le j,k\le 4, \qquad j\ne k,
\end{gather*}
where $L_{jk}=-L_{kj}$. The generators for f\/lat space conformal symmetries are related to these via
\begin{gather*}
P_1= \partial_x=L_{13}+iL_{14},\qquad P_2=\partial_y=L_{23}+iL_{24},\qquad D=iL_{34},\qquad
 J=L_{12},\\
 K_j=L_{j3}-iL_{j4},\qquad j=1,2,\qquad
D=x\partial_x+y\partial_y, \qquad J=x\partial_y-y\partial_x,\\ K_1=2xD-\big(x^2+y^2\big)\partial_x, \qquad \dots.
\end{gather*}

The generators for $2$-sphere conformal constants of the motion are related to the $L_{jk}$ via
\begin{gather*} L_{12}=J_{12}=s_1\partial_{s_2}-s_2\partial_{s_1},\qquad L_{13}=J_{13},\qquad L_{23}=J_{23},\\
 L_{14}=-i\partial_{s_1},\qquad L_{24}=-i\partial_{s_2},\qquad L_{34}=-i\partial_{s_3}.
 \end{gather*}
Note that in identifying tetraspherical coordinates we can always permute the parameters $1$--$4$. Also, we can apply an
arbitrary ${\rm SO}(4,\C)$ transformation to the tetraspherical coordinates, so the above relations between Euclidean and tetraspherical coordinates are far from unique.

 {\bf 2nd order conformal symmetries $\sim H $}:
The 11-dimensional space of conformal symmet\-ries~$\sim H$ has basis
\begin{gather*}
L_{12}^2-L_{34}^2,\quad L_{13}^2-L_{24}^2,\quad L_{23}^2-L_{14}^2,\quad L_{12}^2+L_{13}^2+L_{23}^2,\quad L_{12}L_{34}+L_{23}L_{14}-L_{13}L_{24},\\
\{L_{13},L_{14}\}+\{L_{23},L_{24}\}, \quad \{L_{13},L_{23}\}+\{L_{14},L_{24}\}, \quad \{L_{12},L_{13}\}+\{L_{34},L_{24}\}, \\
 \{L_{12},L_{14}\}-\{L_{34},L_{23}\}, \quad \{L_{12},L_{23}\}-\{L_{34},L_{14}\}, \quad \{L_{12},L_{24}\}+\{L_{34},L_{13}\}.
 \end{gather*}

All of this becomes much clearer if we make use of the decomposition $\mathfrak{so}(4,\C)\equiv \mathfrak{so}(3,\C)\oplus \mathfrak{so}(3,\C)$ and the functional realization of the Lie algebra.
Setting
\begin{gather*} J_1=\tfrac12(L_{23}-L_{14}),\qquad J_2=\tfrac12(L_{13}+L_{24}),\qquad J_3=\tfrac12(L_{12}-L_{34}),\\
 K_1=\tfrac12(L_{23}+L_{14}),\qquad K_2=\tfrac12(L_{13}-L_{24}),\qquad K_3=\tfrac12(L_{12}+L_{34}),
 \end{gather*}
we have $[J_i,J_j]=\epsilon_{ijk}J_k$, $[K_i,K_j]=\epsilon_{ijk}K_k$, $[J_i,K_j]=0$.
In the variables $z=x+iy$, ${\bar z}=x-iy$:
\begin{gather*} J_1=\tfrac12\big(i\partial_z-iz^2\partial_z\big),\qquad J_2=\tfrac12\big(\partial_z+z^2\partial_z\big),\qquad J_3=iz\partial_z,\\
 K_1=\tfrac12\big({-}i\partial_{\bar z}+i{\bar z}^2\partial_{\bar z}\big),\qquad K_2=\tfrac12\big(\partial_{\bar z}+{\bar z}^2\partial_{\bar z}\big),\qquad
K_3=-i{\bar z}\partial_{\bar z},
\end{gather*}
so the $J_i$ operators depend only on $z$ and the $K_j$ operators depend only on $\bar z$. Also
$ J_1^2+J_2^2+J_3^2\equiv 0$, $K_1^2+K_2^2+K_3^2\equiv 0$.
The space of 2nd order elements in the enveloping algebra is thus 21-dimensional and decomposes as $A_z\oplus A_{\bar z}\oplus A_{z{\bar z}}$, where
$A_z$ is 5-dimensional with basis $J_1^2$, $J_3^2$, $\{J_1,J_2\}$, $\{J_1,J_3\}$, $\{J_2,J_3\}$,\
$A_{\bar z}$ is 5-dimensional with basis~$K_1^2$, $K_3^2$, $\{K_1,K_2\}$, $\{K_1,K_3\}$, $\{K_2,K_3\}$,
and $A_{z{\bar z}}$ is 9-dimensional with basis~$J_iK_j$, $1\le i,j\le 3$. Note that all of the elements of $A_{z{\bar z}}$ are~$\sim H$,
whereas none of the nonzero elements of $A_z$, $A_{\bar z}$
have this property. Here, the transposition $J_i\leftrightarrow K_i$ is a conformal equivalence.

\subsection{Classif\/ication of nondegenerate conformally superintegrable systems} \label{3.1}

With this simplif\/ication it becomes feasible to classify all conformally 2nd order superintegrable systems with nondegenerate potential. Since every such system
has generators ${S}^{(1)}={S}_0^{(1)}+W_1(z,{\bar z})$, ${S}^{(2)}={S}_0^{(2)}+W_2(z,{\bar z})$, it is suf\/f\/icient to classify, up to $
{\rm SO}(4,\C)$ conjugacy, all free conformal quadratic algebras
with generators ${S}_0^{(1)}$, ${S}_0^{(2)}$, $\operatorname{mod} {H}_0$ ($H_0=\partial_{z{\bar z}}$) and then to determine for which of these free conformal algebras the
integrability conditions for equations~(\ref{nondegpot1}) hold identically, so that the system admits a nondegenerate potential ${\tilde V}(z,{\bar z})$ which can be computed.
The classif\/ication breaks up into the
following possible cases:
\begin{itemize} \itemsep=0pt
\item Case 1: ${S}_0^{(1)},{S}_0^{(2)}\in A_z$. (This is conformally equivalent to ${S}_0^{(1)},{S}_0^{(2)}\in A_{\bar z}$.)
The possible free conformal quadratic algebras of this type, classif\/ied up to ${\rm SO}(3,\C)$ conjugacy $\operatorname{mod} J_1^2+J_2^2+J_3^2$ can easily be
obtained from the computations in \cite{KM2014}. They are the pairs
\begin{gather*} 1)~J_3^2, \ J_1^2,\qquad 2)~J_3^2, \ \{J_1+iJ_2,J_3\},\qquad 3)~J_3^2,\ \{J_1,J_3\},\\
4)~\{J_2,J_2+iJ_1\},\ \{J_2,J_3\}, \qquad 5)~J_3^2,\ (J_1+iJ_2)^2,\\ 6)~\{J_1+iJ_2,J_3\}, \ (J_1+iJ_2)^2.
\end{gather*}
 Checking pairs $1)$--$5)$ we f\/ind that they do not admit a nonzero potential, so they do not correspond to nondegenerate conformal superintegrable systems. This is in
dramatic distinction to the results of~\cite{KM2014}, where for Helmholtz systems on constant curvature spaces there was a 1-1 relationship between
free quadratic algebras and nondegenerate superintegrable systems. Pair~$6)$, does correspond to a superintegrable system, the exceptional case
${\tilde V}=f(z)$, where~$f(z)$ is arbitrary. (This system is conformally St\"ackel equivalent to the singular Euclidean system $E_{15}$.)
Equivalently, the system in $A_{\bar z}$ with analogous
$K$-operators yields the potential ${\tilde V}=f({\bar z})$, see (\ref{Varb'}) in~Section~\ref{3.2}.
\item Case 2: ${S}_0^{(1)}={S}_J^{(1)}+{S}_K^{(1)}$, ${S}_0^{(2)}={S}_J^{(2)}$, where ${S}_J^{(1)}$, $S_J^{(2)}$ are selected from one of the pairs $1)$--$6)$ above
 and ${S}_K^{(1)}$ is a nonzero element of $A_{\bar z}$. Again there is a conformally equivalent case, where the roles of~$J_i$ and~$K_i$ are switched.
 To determine the possibilities for~${S}_K^{(1)}$ we classify the 2nd order elements in the enveloping algebra of $\mathfrak{so}(3,\C)$ up to ${\rm SO}(3,\C)$ conjugacy,
 $\operatorname{mod} K_1^2+K_2^2+K_3^2$. From the computations in~\cite{KM2014} we see easily that there are the following representatives for the equivalence classes:
 \begin{gather*} a)~K_3^2,\qquad b)~K_1^2+aK_2^2,\quad a\ne 0,1, \qquad c)~(K_1+iK_2)^2,\\
 d)~K_3^2+(K_1+iK_2)^2,\qquad e)~\{K_3,K_1+iK_2\}.
 \end{gather*}
 For pairs 1), 3), 4), 5) above and all choices $a)$--$e)$ we f\/ind that the integrability conditions are never satisf\/ied, so there are no corresponding nondegenerate
 superintegrable systems. For pair $2)$, however, we f\/ind that any choice $a)$--$e)$ leads to the same nondegenerate superintegrable system
 $[2,2]$, see (\ref{V[22norm']}) in Section~\ref{3.2}. While it appears that there are multiple generators for this one system, each set of generators maps to any other
 set by a conformal St\"ackel transformation and a change of variable. For pair $6)$, we f\/ind that any choice $a)$--$e)$ leads to the same nondegenerate superintegrable system~$[4]$, see~(\ref{V[4]norm'}) in Section~\ref{3.2}. Again each set of generators maps to any other
 set by a conformal St\"ackel transformation and a change of variable.

\item Case 3: ${S}_0^{(1)}={S}_J^{(1)}$, ${S}_0^{(2)}={S}_J^{(2)}+{S}_K^{(2)}$, where ${S}_J^{(1)}$, $S_J^{(2)}$
are selected from
one of the pairs $1)$--$6)$ above
 and ${S}_K^{(2)}$ is a nonzero element of~$A_{\bar z}$. Again there is a conformally equivalent case, where
 the roles of~$J_i$ and~$K_i$ are switched.
 To determine the possibilities for ${S}_K^{(2)}$ we classify the 2nd order elements in the enveloping algebra of $\mathfrak{so}(3,\C)$ up to ${\rm SO}(3,\C)$ conjugacy,
 $\operatorname{mod} K_1^2+K_2^2+K_3^2$. They are $a)$--$e)$ above.
 For pairs $1)$--$4)$, $6)$ above and all choices $a)$--$e)$ the integrability conditions are never satisf\/ied, so there are no corresponding nondegenerate
 superintegrable systems. For pair~$5)$, however, we f\/ind that any choice \mbox{$a)$--$e)$} leads to the same nondegenerate superintegrable system
 $[2,2]$, see~(\ref{V[22norm']}) in Section~\ref{3.2}. Again each set of generators maps to any other
 set (and to any $[2,2]$ generators in Case~2) by a conformal St\"ackel transformation and a change of variable.

 \item Case 4: ${S}_0^{(1)}={S}_J^{(1)}$, ${S}_0^{(2)}={S}_K^{(2)}$, where ${S}_J^{(1)}$ is selected from
one of the representatives \mbox{$a)$--$e)$} above and ${S}_K^{(2)}$ is selected from one of the analogous representatives $a)$--$e)$ expressed as $K$-operators.
We f\/ind that each of the~25 sets of generators leads to the single conformally superintegrable system~$[0]$, see (\ref{V[0]norm'}) in Section~\ref{3.2}, and each set of generators
maps to any other
 set by a conformal St\"ackel transformation and a change of variable.
 \item Case 5: ${S}_0^{(1)}={S}_J^{(1)}+{S}_K^{(1)}$, ${S}_0^{(2)}={S}_J^{(2)}+{S}_K^{(2)}$, where ${S}_J^{(1)}$, $S_J^{(2)}$ are selected from
one of the pairs $1)$--$6)$ above and ${S}_K^{(1)}$, $S_K^{(2)}$ are obtained from ${S}_J^{(1)}$, $S_J^{(2)}$, respectively, by replacing each~$J_i$ by~$K_i$. We f\/ind the following possibilities:
\begin{enumerate}\itemsep=0pt
 \item[i)] ${S}_0^{(1)}=J_1^2+K_1^2$, ${S}_0^{(2)}=J_3^2+K_3^2$. This extends to the system $[1,1,1,1]$, see (\ref{V[1111norm']}) in Section~\ref{3.2}.
 \item[ii)] ${S}_0^{(1)}=J_3^2+K_3^2$, ${S}_0^{(2)}=\{J_3,J_1+iJ_2\}+\{K_3,K_1+iK_2\}$. This extends to the system $[2,1,1]$, see (\ref{V211norm'}) in Section~\ref{3.2}.
 \item[iii)] ${S}_0^{(1)}=J_3^2+K_3^2$, ${S}_0^{(2)}=\{J_1,J_3\}+\{K_1,K_3\}$. This extends to the system $[1,1,1,1]$, see (\ref{V[1111norm']}) in Section~\ref{3.2}, again,
 equivalent to the generators $i)$ by a conformal St\"ackel transformation and a change of variable.
 \item[iv)] ${S}_0^{(1)}=\{J_1,J_2+iJ_1\}+\{K_1,K_2+iK_1\}$, ${S}_0^{(2)}=\{J_2,J_3\}+\{K_2,K_3\}$.
 This does not extend to a conformal superintegrable system.
\item[v)] ${S}_0^{(1)}=(J_1+iJ_2)^2+(K_1+iK_2)^2$, ${S}_0^{(2)}=J_3^2+K_3^2$. This extends to the system $[2,1,1]$, see~(\ref{V211norm'}) in Section~\ref{3.2}, again,
 equivalent to the generators $ii)$ by a conformal St\"ackel transformation and a change of variable.
 \item[vi)] ${S}_0^{(1)}=\{J_3,J_1+iJ_2\}+\{K_3,K_1+iK_2\}$, ${S}_0^{(2)}=(J_1+iJ_2)^2+(K_1+iK_2)^2$,
 which extends to the system with potential~$[3,1]$, see (\ref{V[31]norm'}) in Section~\ref{3.2}.
\end{enumerate}
\end{itemize}

\begin{Example}
 We describe how apparently distinct superintegrable systems of a f\/ixed type are actually the same.
 In Case 2 consider the system with generators $\{J_1+iJ_2,J_3\}+(K_1+iK_2)^2$, $(J_1+iJ_2)^2$. This extends to the conformally superintegrable system
 $[4]$ with
 f\/lat space Hamiltonian operator $H_1=\partial_{z{\bar z}}+ V^{(1)}$, where
 $V^{(1)}=2k_3z{\bar z}+2k_4z+k_3{\bar z}^3+3k_4{\bar z}^2+k_1{\bar z}+k_2$. The system with generators
 $\{J_1+iJ_2,J_3\}+K_3^2+(K_1+iK_2)^2$, $(J_1+iJ_2)^2$ again extends to the conformally superintegrable system~$[4]$. Indeed, replacing $z$, ${\bar z}$ by
 $Z$, ${\bar Z}$ to distinguish the
 two systems, we f\/ind the 2nd
 f\/lat space Hamiltonian operator $H_2=\partial_{Z{\bar Z}}+ V^{(2)}$, where
 \begin{gather*}
 V^{(2)}=\frac{c_3\operatorname{arcsinh}^3({\bar Z})+3c_4\operatorname{arcsinh}^2({\bar Z})+(2c_3 Z+c_1)\operatorname{arcsinh}({\bar Z})+2c_4 Z+c_2}{\sqrt{1-{\bar Z}^2}}.\end{gather*}
 Now we perform a conformal St\"ackel transform on $H_2$ to obtain the new f\/lat space system
 \begin{gather*}
 {\tilde H}_2=\sqrt{1-{\bar Z}^2}\partial_{Z{\bar Z}}+
 c_3\operatorname{arcsinh}^3({\bar Z})+3c_4\operatorname{arcsinh}^2({\bar Z})\\
 \hphantom{{\tilde H}_2=}{} +(2c_3 Z+c_1)\operatorname{arcsinh}({\bar Z})+2c_4 Z+c_2.
 \end{gather*}
 Making the change of variable ${\bar Z}=\sinh W $, we f\/ind
\begin{gather*}
 {\tilde H}_2= \partial_{ZW}+ c_3 W^3+3c_4W^2+(2c_3 Z+c_1)W+2c_4 Z+c_2.
\end{gather*}
 Thus, with the identif\/ications $Z=z$, $W={\bar z}$, $c_i=k_i$, we see that $H_1\equiv {\tilde H}_2$.
\end{Example}

This completes the classif\/ication. The results are summarized in the next section.

\subsection{The 8 Laplace superintegrable systems with nondegenerate potentials}\label{3.2}
The systems are all of the form
\begin{gather*}
\left(\sum_{j=1}^4\partial_{x_j}^2+V({\bf x})\right)\Psi=0, \qquad \text{or}\qquad \big(\partial_x^2+\partial_y^2+{\tilde V}\big)\Psi=0
\end{gather*}
as a f\/lat space system in Cartesian coordinates. The potentials are
\begin{gather}
V_{[1,1,1,1]}=\frac{a_1}{x_1^2}+\frac{a_2}{x_2^2}+\frac{a_3}{x_3^2}+\frac{a_4}{x_4^2},\nonumber\\
{\tilde V}_{[1,1,1,1]}=\frac{a_1}{x^2}+\frac{a_2}{y^2}+\frac{4a_3}{(x^2+y^2-1)^2}-\frac{4a_4}{(x^2+y^2+1)^2},\label{V[1111norm']}
\\
V_{[2,1,1]}=\frac{a_1}{x_1^2}+\frac{a_2}{x_2^2}+\frac{a_3(x_3-ix_4)}{(x_3+ix_4)^3}+\frac{a_4}{(x_3+ix_4)^2},\nonumber\\
{\tilde V}_{[2,1,1]}=\frac{a_1}{x^2}+\frac{a_2}{y^2}-a_3\big(x^2+y^2\big)+a_4, \label{V211norm'}\\
V_{[2,2]}=\frac{a_1}{(x_1+ix_2)^2}+\frac{a_2(x_1-ix_2)}{(x_1+ix_2)^3}
+\frac{a_3}{(x_3+ix_4)^2}+\frac{a_4(x_3-ix_4)}{(x_3+ix_4)^3},\nonumber\\
{\tilde V}_{[2,2]}=\frac{a_1}{(x+iy)^2}+\frac{a_2(x-iy)}{(x+iy)^3}
+a_3-a_4\big(x^2+y^2\big),\label{V[22norm']}\\
V_{[3,1]}=\frac{a_1}{(x_3+ix_4)^2}+\frac{a_2x_1}{(x_3+ix_4)^3}
+\frac{a_3(4{x_1}^2+{x_2}^2)}{(x_3+ix_4)^4}+\frac{a_4}{{x_2}^2},\nonumber\\
 {\tilde V}_{[3,1]}=a_1-a_2x
+a_3\big(4x^2+{y}^2\big)+\frac{a_4}{{y}^2},\label{V[31]norm'} \\
V_{[4]}=\frac{a_1}{(x_3+ix_4)^2}+a_2\frac{x_1+ix_2}{(x_3+ix_4)^3}
+a_3\frac{3(x_1+ix_2)^2-2(x_3+ix_4)(x_1-ix_2)}{(x_3+ix_4)^4}\nonumber\\
\hphantom{V_{[4]}=}{}
 +a_4\frac{4(x_3+ix_4)(x_3^2+x_4^2)+2(x_1+ix_2)^3}{(x_3+ix_4)^5},\nonumber\\
 {\tilde V}_{[4]}=a_1-a_2(x+iy)
+a_3\big(3(x+iy)^2+2(x-iy)\big)
-a_4\big(4\big(x^2+y^2\big)+2(x+iy)^3\big),\label{V[4]norm'}\\
V_{[0]}=\frac{a_1}{(x_3+ix_4)^2}+\frac{a_2x_1+a_3x_2}{(x_3+ix_4)^3}+a_4\frac{x_1^2+x_2^2}{(x_3+ix_4)^4},\nonumber\\
 {\tilde V}_{[0]}=a_1-(a_2x+a_3y)+a_4\big(x^2+y^2\big),\label{V[0]norm'}\\
V_{\rm arb}=\frac{1}{(x_3+ix_4)^2}f\left(\frac{-x_1-ix_2}{x_3+ix_4}\right),\nonumber\\
 {\tilde V}_{\rm arb}=f({x+iy}), \qquad \text{$f$ is arbitrary},\label{Varb'}\\
V(1)=a_1\frac{1}{(x_1+ix_2)^2}+a_2\frac{1}{(x_3+ix_4)^2}
+a_3\frac{(x_3+ix_4)}{(x_1+ix_2)^3}+a_4\frac{(x_3+ix_4)^2}{(x_1+ix_2)^4},\nonumber\\
 {\tilde V}(1)=\frac{a_1}{(x+iy)^2}+a_2
-\frac{a_3}{(x+iy)^3}+\frac{a_4}{(x+iy)^4} \qquad (\text{a special case of (\ref{Varb'})}),\label{V[1]norm'}
\\
V(2)'=a_1\frac{1}{(x_3+ix_4)^2}+a_2\frac{(x_1+ix_2)}{(x_3+ix_4)^3}
+a_3\frac{(x_1+ix_2)^2}{(x_3+ix_4)^4}+a_4\frac{(x_1+ix_2)^3}{(x_3+ix_4)^5},\nonumber\\
 {\tilde V}(2)'=a_1+a_2(x+iy)
+a_3(x+iy)^2+a_4(x+iy)^3 \qquad (\text{a special case of (\ref{Varb'})}).\label{V[2]norm'}
\end{gather}
We note that systems (\ref{V[1]norm'}), (\ref{V[2]norm'}) are not the fundamental B\^ocher classes; they are
merely special cases of the singular system~(\ref{Varb'}). We list them because they, and not the general~(\ref{Varb'}),
appear as contractions of the fundamental systems.

\subsection[Contractions of conformal superintegrable systems with potential induced by generalized In\"on\"u--Wigner contractions]{Contractions of conformal superintegrable systems\\ with potential induced by generalized In\"on\"u--Wigner contractions}\label{3.3}

\looseness=1
The basis symmetries ${\cal S}^{(j)} ={\cal S}^{(j)}_0+W^{(j)}$, ${\cal H}={\cal H}_0+V$ of a nondegenerate 2nd order conformally
superintegrable system determine a conformal quadratic algebra~(\ref{confquadalg}), and if the parameters of the potential are set equal to~$0$, the free system ${\cal S}^{(j)}_0$, ${\cal H}_0$, $j=1,2$ also determines a conformal
quadratic algebra without parameters, which we call a {\it free conformal quadratic algebra}. The elements of this free algebra belong
to the enveloping algebra of~$\mathfrak{so}(4,\C)$
with basis~(\ref{conformalKV}). Since the system is nondegenerate the integrability conditions for the potential are satisf\/ied identically and the
full quadratic algebra can be computed from the free algebra,
modulo a choice of basis for the 4-dimensional potential space. Once we choose a basis for~$\mathfrak{so}(4,\C)$, its enveloping algebra is uniquely
determined by the structure constants. Structure relations in the enveloping
algebra are continuous functions of the structure constants, so a contraction of one $\mathfrak{so}(4,\C)$ to itself induces a contraction of the
 enveloping algebras. Then the free conformal quadratic algebra constructed in the enveloping algebra will contract
to another free quadratic algebra. (In~\cite{KM2014} essentially the same argument was given in more detail for Helmholtz superintegrable
systems on constant curvature spaces.)

In this paper we consider a family of contractions of $\mathfrak{so}(4,\C)$ to itself that we call B\^ocher contractions. All these contractions are implemented via coordinate
transformations. Suppose we have a conformal nondegenerate superintegrable system with free generators ${\cal H}_0$, ${\cal S}^{(1)}_0$, ${\cal S}^{(2)}_0$
that determines the conformal and free conformal quadratic algebras $Q$ and $Q^{(0)} $
and has structure functions $A^{ij}({\bf x})$, $B^{ij}({\bf x})$, $C^{ij}({\bf x})$ in Cartesian coordinates ${\bf x}=(x,y)$.
Further, suppose this system contracts to another nondegenerate system
${\cal H'}_0$, ${\cal S'}^{(1)}_0$, ${\cal S'}^{(2)}_0$ with conformal quadratic algebra~${Q'}^{(0)}$. We show here that this contraction
induces a contraction of the associated nondegenerate superintegrable system
${\cal H}={\cal H}_0+V$, ${\cal S}^{(1)}={\cal L}^{(1)}_0+W^{(1)}$,
 ${\cal S}^{(2)}={\cal S}^{(2)}_0+W^{(2)}$, $Q$ to
${\cal H}'={{\cal H}'}_0+V'$, ${\cal S'}^{(1)}={\cal S'}_0^{(1)}+{W^{(1)}}'$,
 ${\cal S'}^{(2)}={\cal S'}_0^{(2)}+{W^{(2)}}'$,~$Q'$.
The point is that in the contraction process the symmetries ${{\cal H}'}_0(\epsilon)$,
${\cal S'}^{(1)}_0(\epsilon)$,
$ {\cal S'}^{(2)}_0(\epsilon)$
remain continuous functions of~$\epsilon$, linearly independent as quadratic forms, and
 $\lim\limits_{\epsilon\to 0} {\cal H'}_0(\epsilon)={{\cal H'}}_0$,
$\lim\limits_{\epsilon\to 0} {\cal S'}^{(j)}_0(\epsilon)={\cal S'}_0^{(j)}$.
Thus the asso\-cia\-ted functions $A^{ij}(\epsilon)$, $B^{ij}(\epsilon)$, $C^{(ij)}$ will also be continuous functions of $\epsilon$ and
$\lim\limits_{\epsilon\to 0}A^{ij}(\epsilon)={A'}^{ij}$, $\lim\limits_{\epsilon\to 0}B^{ij}(\epsilon)={B'}^{ij}$, $\lim\limits_{\epsilon\to 0}C^{ij}(\epsilon)={C'}^{ij}$.
Similarly, the integrability conditions for the potential equations
\begin{gather*}
 V^{(\epsilon)}_{22} = V^{(\epsilon)}_{11} + A^{22}(\epsilon) V^{(\epsilon)}_1+B^{22}(\epsilon) V^{(\epsilon)}_2+C^{22}(\epsilon) V^{(\epsilon)},\\
 V^{(\epsilon)}_{12} = A^{12}(\epsilon) V^{(\epsilon)}_1+B^{12}(\epsilon) V^{(\epsilon)}_2+C^{12}(\epsilon) V^{(\epsilon)},
\end{gather*}
will hold for each $\epsilon$ and in the limit.
This means that the 4-dimensional solution space for the potentials $V$ will deform continuously into the 4-dimensional solution space for
the potentials~$V'$. Thus the target space of solutions~$V'$ (and of the functions~$W'$) is uniquely determined by the free quadratic algebra contraction.

There is an apparent lack of uniqueness in this procedure, since for a nondegenerate superintegrable system one typically
chooses a basis $V^{(j)}$, $j=1,\dots,4$ for the potential space and expresses a general potential as $V=\sum\limits_{j=1}^4a_jV^{(j)}$.
Of course the choice of basis for the source system is arbitrary, as is the choice for the target system.
Thus the structure equations for the quadratic algebras and the dependence~$a_j(\epsilon)$ of the contraction constants
on~$\epsilon$ will vary depending on these choices. However, all such possibilities are related by a basis change matrix.

\subsection{Relation to separation of variables and B\^ocher's limit procedures} \label{3.4}

B\^ocher's analysis \cite{Bocher,Proceedings} involves symbols of the form $[n_1,n_2,\dots,n_p]$, where
$n_1+\dots +n_p=4$. These symbols are used to def\/ine coordinate surfaces as
follows. Consider the quadratic forms
\begin{gather} \label{ellipsoidalcoords}\Omega =x^2_1+x^2_2+x^2_3+x^2_4=0,\qquad
\Phi =\frac{x^2_1}{\lambda -e_1} + \frac{x^2_2}{\lambda -e_2} +
\frac{x^2_3}{\lambda -e_3} + \frac{x^2_4}{\lambda -e_4}=0.
\end{gather}
If $e_1,e_2,e_3,e_4$ are pairwise distinct, the elementary divisors of these two forms are denoted by the symbol
$[1,1,1,1]$, see~\cite{Bromwich}. Given a point in 2D f\/lat space with Cartesian coordinates~$(x^0,y^0)$, there corresponds a set of tetraspherical coordinate
$(x^0_1,x^0_2,x^0_3,x^0_4)$, unique up to multiplication by a nonzero constant. If we substitute these coordinates into expressions~(\ref{ellipsoidalcoords})
we can verify that there are exactly 2 roots $\lambda=\rho,\mu$ such that $\Phi=0$.
These are elliptic coordinates. It can be verif\/ied that they are orthogonal with respect to the metric $ds^2=dx^2+dy^2$ and that
they are $R$-separable for the
Laplace equations $(\partial^2_x+\partial^2_y)\Theta=0$ or $\Big(\sum\limits_{j=1}^4\partial_{x_j}^2\Big)\Theta=0$.
Now consider the potential
$V_{[1,1,1,1]}=\frac{a_1}{x^2_1} + \frac{a_2}{x^2_2} +
\frac{a_3}{x^2_3} + \frac{a_4}{x^2_4}$.
It can be verif\/ied that this is the only possible potential $V$ such that the Laplace equation $\Big(\sum\limits_{j=1}^4\partial_{x_j}^2+V\Big)\Theta=0$ is $R$-separable in elliptic
coordinates for {\it all} choices of the parameters~$e_j$. The separation is characterized by 2nd order conformal symmetry operators that are
linear in the parameters~$e_j$. In particular the symmetries span a 3-dimensional subspace of symmetries as the~$e_j$ are varied, so the system $\Big(\sum\limits_{j=1}^4\partial_{x_j}^2+V_{[1,1,1,1]}\Big)\Theta=0$ must be conformally
superintegrable. We can write this as
\begin{gather*}H=(x_3+ix_4)^2\left(\partial ^2_{x_1}+ \partial ^2_{x_2}+ \partial ^2_{x_3}+
\partial ^2_{x_4} +\frac{a_1}{x^2_1} +\frac {a_2}{x^2_2} + \frac{a_3}{x^2_3} +
\frac{a_4}{x^2_4}\right),\end{gather*}
or in terms of f\/lat space coordinates $x$, $y$ as
\begin{gather*}
H= \partial_x^2+\partial_y^2+\frac{a_1}{x^2}+\frac{a_2}{y^2}+\frac{4a_3}{(x^2+y^2-1)^2}-\frac{4a_4}{(x^2+y^2+1)^2}.
\end{gather*}
For the coordinates $s_i$, $i=1,2,3$ we obtain
\begin{gather*}
H=(1+s_3)^2\left(\partial ^2_{s_1}+\partial ^2_{s_2}+\partial ^2_{s_3}
-\frac{a_1}{s^2_1} - \frac{a_2}{s^2_2} - \frac{a_3}{s^2_3} -a_4\right).
\end{gather*}
The coordinate curves are described by $[1,1,1,\overset{\infty}{1}
]$ (because we can always transform to equivalent coordinates for which $e_4=\infty$) and the corresponding
$H\Theta=0$ system is proportional to~$S_9$, the eigenvalue equation for the generic potential on the 2-sphere,
which separates variables in elliptic coordinates
$s^2_i=\frac{(\rho -e_i)(\mu -e_i)}{(e_i-e_j)(e_i-e_k)}$,
where $(e_i-e_j)(e_i-e_k)\neq 0$ and $i,j,k=1,2,3$.
The quantum Hamiltonian when written using these coordinates is equivalent to
\begin{gather*}
{\cal H}=\frac{1}{\rho -\mu}\left[P_\rho^2-P_\mu^2 -\sum ^3_{i=1}
a_i\frac{(e_i-e_j)(e_i-e_k)}{(\rho -e_i)(\mu -e_i)}\right],
\end{gather*}
where $P_\lambda=\sqrt{\prod\limits^3_{i=1}(\lambda -e_i)}\partial_\lambda$.

\subsection[$\protect{[1,1,1,1]}$ to $\protect{[2,1,1]}$ contraction]{$\boldsymbol{[1,1,1,1]}$ to $\boldsymbol{[2,1,1]}$ contraction}

B\^ocher provides a recipe to derive separable coordinates in the cases, where some of the $e_i$ become equal. In particular,
B\^ocher shows that the process of
making $e_1\rightarrow e_2$ together with suitable transformations of the
$a_i's$ produces a conformally equivalent $H$. This corresponds to the choice of
coordinate curves obtained by the B\^ocher limiting process $[1,1,1,1]\to [2,1,1]$, i.e.,
$ e_1=e_2+\epsilon ^2$,
$x_1\rightarrow \frac{iy_1}{\epsilon}$,
$x_2\rightarrow \frac{y_1}{\epsilon} + \epsilon y_2$,
$x_j\rightarrow y_j$, $j=3,4$,
which results in the pair of quadratic forms
\begin{gather*}\Omega =2y_1y_2+y^2_3+y^2_4=0,\qquad \Phi =\frac{y^2_1}{(\lambda -e_2)^2}+\frac{2y_1y_2}{(\lambda -e_2)} +
\frac{y^2_3}{(\lambda -e_3)} +\frac {y^2_4}{(\lambda -e_4)} =0.
\end{gather*}
The coordinate curves with $e_4= \infty$ correspond to cyclides with
elementary divisors $[2,1,\overset{\infty}{1} ]$, see~\cite{Bromwich}, i.e.,
\begin{gather*}
\Phi =\frac{y^2_1}{(\lambda -e_2)^2}+\frac{2y_1y_2}{(\lambda -e_2)} +
\frac{y^2_3}{(\lambda -e_3)}=0.
\end{gather*}

The $\lambda$ roots of $\Phi$ yield planar elliptic coordinates.
In order to identify ``Cartesian'' coordinates on the cone we can choose
$ y_1=\frac{1}{\sqrt{2}}(x'_1+ix'_2)$,
$y_2=\frac{1}{\sqrt{2}}(x'_1-ix'_2)$, $y_3=x'_3$, $y_4=x'_4$.
Note that the composite linear coordinate mapping
\begin{gather}
x_1+ix_2=\frac{i\sqrt{2}}{\epsilon}(x'_1+ix'_2)+\frac{i\epsilon}{\sqrt{2}}(x'_1-ix'_2) , \qquad
x_1-ix_2=-\frac{i\epsilon}{\sqrt{2}}(x'_1-ix'_2),\nonumber\\
x_3=x'_3,\qquad x_4=x'_4,\label{coordlimit}
\end{gather}
satisf\/ies $\lim\limits_{\epsilon\to 0} \sum\limits_{j=1}^4 x_j^2= \sum\limits_{j=1}^4 {x'}_j^2=0$, preserving the null cone, and it induces a~contraction of the
Lie algebra $\mathfrak{so}(4,\C)$ to itself. An explicit computation yields the B\^ocher contraction $[1,1,1,1]\to [2,1,1]$:
 \begin{gather*} L'_{12}=L_{12},\qquad L'_{13}=-\frac{i}{\sqrt{2}\epsilon}(L_{13}-iL_{23})-\frac{i\epsilon}{\sqrt{2}}L_{13},\\
 L'_{23}=-\frac{i}{\sqrt{2}\epsilon}(L_{13}-iL_{23})-\frac{\epsilon}{\sqrt{2}}L_{13},\qquad
 L'_{34}=L_{34}, \\
 L'_{14}=-\frac{i}{\sqrt{2}\epsilon}(L_{14}-iL_{24})-\frac{i\epsilon}{\sqrt{2}}L_{14},\qquad
 L'_{24}=-\frac{i}{\sqrt{2}\epsilon}(L_{14}-iL_{24})-\frac{\epsilon}{\sqrt{2}}L_{14}.
 \end{gather*}

Now under the contraction $[1,1,1,1]\to [2,1,1]$ we have
$ V_{[1,1,1,1]}
\overset{\epsilon\to0}{\Longrightarrow}V_{[2,1,1]}$, where
\begin{gather}
 V_{[2,1,1]}=\frac{b_1}{(x'_1+ix'_2)^2}+\frac{b_2(x'_1-ix'_2)}{(x'_1+ix'_2)^3}+\frac{b_3}{{x'_3}^2}+\frac{b_4}{{x'_4}^2},\nonumber\\
 a_1=-\frac12\left(\frac{b_1}{\epsilon^2}+\frac{b_2}{2\epsilon^4}\right),\qquad a_2=- \frac{b_2}{4\epsilon^4},
 \qquad a_3=b_3,\qquad a_4=b_4.\label{V[211]}
\end{gather}

{\bf Basis of conformal symmetries for original system}:
Let $H_0=\sum\limits_{j=1}^4\partial_{x_j}^2$. A basis is
$ \{H_0+V_{[1,1,1,1]},Q_{12},Q_{13}\}$,
where
$Q_{jk}=L_{jk}^2+a_j\frac{x_k^2}{x_j^2}+a_k\frac{x_j^2}{x_k^2}$, $1\le j<k\le 4$.

{\bf Contraction of basis}: Using the notation of (\ref{V[211]}), we have
\begin{gather*}
 H_0 + V_{[1,1,1,1]}\to H'_0+V_{[2,1,1]},\\
 Q'_{12}=Q_{12}-\frac{b_1}{2\epsilon^2}-\frac{b_2}{2\epsilon^4}
 =(L'_{12})^2+b_1(\frac{x_1'-ix_2'}{x_1'+ix_2'})+b_2\left(\frac{x_1'-ix_2'}{x_1'+ix_2'}\right)^2,\\
 Q'_{13} = 2\epsilon^2 Q_{13}=(L'_{23}-iL'_{13})^2+\frac{b_2{x'_3}^2}{(x'_1+ix'_2)^2}-\frac{b_3(x'_1+ix'_2)^2}{{x'_3}^2}.
\end{gather*}
If we apply the same $[1,1,1,1]\to [2,1,1]$ contraction to the $[2,1,1]$ system, the system contracts to itself, but with parameters
$c_1,\dots,c_4$, where $b_1=-\frac{2c_1}{\epsilon^2}$, $b_2=\frac{c_1}{\epsilon^2}+\frac{4c_2}{\epsilon^4}$, $b_3=c_3$, $b_4=c_4$.

If we apply the same contraction to the $[2,2]$ system, the system contracts to itself, but with altered parameters.
If we apply the same contraction to the $[3,1]$ system, the system contracts to~$V(1)$. If we apply the same contraction to the~$[4]$ system the system contracts to a system with potential
\begin{gather*}
V[0]= \frac{c_1}{(x'_1+ix'_2)^2}+\frac{c_2x'_3+c_3x'_4}{(x'_1+ix'_2)^3}+c_4\frac{{x'}_3^2+{x'}_4^2}{(x'_1+ix'_2)^4}.
\end{gather*}
If we apply this same contraction to the $[0]$, $(1)$ and $(2)$ systems they contract to themselves, but with altered parameters.

The remaining contractions are derived from the B\^ocher recipe~\cite{Bocher,Proceedings}.

\subsection[$\protect{[1,1,1,1]}$ to $\protect{[2,2]}$ contraction]{$\boldsymbol{[1,1,1,1]}$ to $\boldsymbol{[2,2]}$ contraction}

\vspace{-5mm}

\begin{gather*} L'_{12}=L_{12},\qquad L'_{34}=L_{34},\qquad L'_{24}+L'_{13}=L_{24}+L_{13},\\
 L'_{24}-L'_{13}=\left(\epsilon^2+\frac{1}{\epsilon^2}\right)L_{13}-\frac{1}{\epsilon^2}(iL_{14}-L_{24}-iL_{23}),\\
 L'_{23}-L'_{14}=2L_{23}+iL_{13}-iL_{24},\\
 L'_{23}+L'_{14}=i\left(\left(\epsilon^2-\frac{1}{\epsilon^2}\right)L_{13}+\frac{1}{\epsilon^2}(iL_{14}+L_{24}+iL_{23})\right).
 \end{gather*}

 {\bf Coordinate implementation}:
\begin{gather*} x_1=\frac{i}{\sqrt{2}\epsilon}(x'_1+ix'_2),\qquad x_2=\frac{1}{\sqrt{2}}\left(\frac{x'_1+ix'_2}{\epsilon}+\epsilon
(x'_1-ix'_2)\right),\\
 x_3=\frac{i}{\sqrt{2}\epsilon}(x'_3+ix'_4),\qquad x_4=\frac{1}{\sqrt{2}}\left(\frac{x'_3+ix'_4}{\epsilon}+\epsilon
(x'_3-ix'_4)\right).
\end{gather*}

{\bf Limit of 2D potential}: $V_{[1,1,1,1]} \overset{\epsilon\to0}{\Longrightarrow}V_{[2,2]}$,
where
\begin{gather*}
V_{[2,2]}=\frac{b_1}{(x'_1+ix'_2)^2}+\frac{b_2(x'_1-ix'_2)}{(x'_1+ix'_2)^3}
+\frac{b_3}{(x'_3+ix'_4)^2}+\frac{b_4(x'_3-ix'_4)}{(x'_3+ix'_4)^3},
\end{gather*}
and
$ a_1=-\frac12\frac{b_1}{\epsilon^2}-\frac{b_2}{4\epsilon^4}$, $a_2=- \frac{b_2}{4\epsilon^4}$,
$a_3=-\frac12\frac{b_3}{\epsilon^2}-\frac{b_4}{4\epsilon^4}$,
$a_4=- \frac{b_4}{4\epsilon^4}$.

{\bf Contracted basis}:
\begin{gather*}Q_{12}-\frac{b_2}{2\epsilon^4}-\frac{b_1}{2\epsilon^2}\to Q_1'={ L'}_{12}^2+b_1\frac{x'_1-ix'_2}{x'_1+ix'_2}+b_2\frac{(x'_1-ix'_2)^2}{(x'_1+ix'_2)^2},\\
 4\epsilon^4 Q_{13}\to Q_2'=(L'_{13}+iL'_{14}+iL'_{23}-L'_{24})^2-b_2\frac{(x'_3+ix'_4)^2}{(x'_1+ix'_2)^2}-b_4\frac{(x'_1+ix'_2)^2}{(x'_3+ix'_4)^2}.
 \end{gather*}

\subsection[$\protect{[2,1,1]}$ to $\protect{[3,1]}$ contraction]{$\boldsymbol{[2,1,1]}$ to $\boldsymbol{[3,1]}$ contraction}

\vspace{-5mm}

\begin{gather*} L'_{24}=\frac{\sqrt{2}i}{2\epsilon}(L_{14}+iL_{24})-L_{34},L'_{14}+iL'_{34}=-i\epsilon(L_{14}+iL_{24}),\\
L'_{14}-iL'_{34}=\frac{1}{\epsilon}\left(iL_{14}\left(1+\frac{1}{2\epsilon^2}\right)
+L_{24}(1-\frac{1}{2\epsilon^2})-\frac{\sqrt{2}}{\epsilon}L_{34}\right),\\
L'_{13}=-L_{12}-2\sqrt{2}L_{13}\big(\epsilon+2\epsilon^3\big),\qquad
 L'_{23}+iL'_{12}=4\epsilon^3 L_{13}, \\
 L'_{23}-iL'_{12}=\left(2\sqrt{2}-\frac{\sqrt{2}}{\epsilon^2}\right)L_{12}+\left(8\epsilon^3+4\epsilon-\frac{2}{\epsilon}
 +\frac{1}{2\epsilon^3}\right)L_{13}+\frac{i}{2\epsilon^3}L_{23}.
 \end{gather*}

{\bf Coordinate implementation}:
\begin{gather*} x_1+ix_2=-\frac{i\sqrt{2}\epsilon}{2}x_2' +\frac{(i x'_1- x'_3)}{\epsilon},\qquad\!
x_1-ix_2=-\epsilon (x_3'+ix'_1)
+\frac{3i\sqrt{2}x'_2}{4\epsilon}+\frac12 \frac{(ix'_1- x'_3)}{\epsilon^3},\\
 x_3=-\frac12 x'_2-\frac{\sqrt{2}}{2} \frac{(x'_1+i x'_3)}{\epsilon^2},\qquad x_4=x_4'.
 \end{gather*}

{\bf Limit of 2D potential}: $V_{[2,1,1]}
\overset{\epsilon\to0}{\Longrightarrow}V_{[3,1]}$,
where
\begin{gather}\label{V[31]} V_{[3,1]}=\frac{c_1}{(x'_1+ix'_3)^2}+\frac{c_2x'_2}{(x'_1+ix'_3)^3}
+\frac{c_3(4{x'_2}^2+{x'_4}^2)}{(x'_1+ix'_3)^4}+\frac{c_4}{{x'_4}^2},\\
 b_1=\frac{c_3}{\epsilon^6}+\frac{\sqrt{2}c_2}{4\epsilon^4}-\frac{c_1}{\epsilon^2},\qquad b_2=-
\frac{c_3}{\epsilon^4}-\frac{\sqrt{2}c_2}{2\epsilon^2},\qquad b_3=\frac{c_3}{4\epsilon^8},\qquad b_4=c_4.\nonumber
\end{gather}

 {\bf Basis of conformal symmetries for original system $H_0+V_{[2,1,1]}$}:
\begin{gather*} Q_{12}=(L_{12})^2+b_1\left(\frac{x_1-ix_2}{x_1+ix_2}\right)+b_2\left(\frac{x_1-ix_2}{x_1+ix_2}\right)^2,\\
Q_{13}=(L_{23}-iL_{13})^2+\frac{b_2{x_3}^2}{(x_1+ix_2)^2}-\frac{b_3(x_1+ix_2)^2}{{x_3}^2}.
\end{gather*}

 {\bf Contraction of basis}:
\begin{gather*} H_0+V_{[2,1,1]}\to H_0'+V_{[3,1]},\\
 Q'_{12}=-2\epsilon^4Q_{12}+\frac{c_3}{2\epsilon^4}-c_1=(L'_{12}-iL'_{23})^2+\frac{c_2x_2'}{x_1'+ix_3'}
+\frac{4c_3{x_2'}^2}{(x_1'+ix_3')^2},\\
 Q'_{13}=-\frac{\sqrt{2}}{4}\left(Q_{13}+2\epsilon^2 Q_{12}-\frac{3c_3}{2\epsilon^6}-\frac{\sqrt{2}c_2}{4\epsilon^4}+c_1\right)\\
\hphantom{Q'_{13}}{} =\frac12\{L'_{13},L'_{23}+iL'_{12}\}
+\frac{c_1x'_2}{x'_1+ix'_3}+\frac{c_2({x'_4}^2+4{x'_2}^2)}{4(x'_1+ix'_3)^2}+\frac{2 c_3x'_2({x'_4}^2+2{x'_2}^2)}{(x'_1+ix'_3)^3}.
\end{gather*}

\subsection[$\protect{[1,1,1,1]}$ to $\protect{[4]}$ contraction]{$\boldsymbol{[1,1,1,1]}$ to $\boldsymbol{[4]}$ contraction}

In this case there is a 2-parameter family of contractions, but all lead to the same result.
Let~$A$,~$B$ be constants such that $AB(1-A)(1-B)(A-B)\ne 0$.

{\bf Coordinate implementation:}
\begin{gather*} x_1=\frac{i}{\sqrt{2AB}\epsilon^3}(x'_1+ix'_2), \\ x_2=\frac{(x'_1+ix'_2)+\epsilon^2(x'_3+ix'_4)+\epsilon^4(x'_3-ix'_4)+\epsilon^6(x'_1-ix'_2)}{\sqrt{2(A-1)(B-1)}\epsilon^3},\\
x_3=\frac{(x'_1+ix'_2)+A\epsilon^2(x'_3+ix'_4)+A^2\epsilon^4(x'_3-ix'_4)+A^3\epsilon^6(x'_1-ix'_2)}{\sqrt{2A(A-1)(A-B)}\epsilon^3},\\
x_4=\frac{(x'_1+ix'_2)+B\epsilon^2(x'_3+ix'_4)+B^2\epsilon^4(x'_3-ix'_4)+B^3\epsilon^6(x'_1-ix'_2)}{\sqrt{2B(B-1)(B-A)}\epsilon^3},
\\
iL'_{14}+iL'_{23}+L'_{13}-L'_{24} = -2i\epsilon^4\sqrt{AB(A-1)(B-1)}L_{12},\\
iL'_{14}-iL'_{23}-L'_{13}-L'_{24} = 2i\epsilon^2\left(\sqrt{B(A-1)(A-B)}L_{13}-\sqrt{AB(A-1)(B-1)}L_{12}\right),\nonumber\\
L'_{12} = \frac{\sqrt{AB}}{\sqrt{(A-1)(B-1)}}L_{12}+\frac{\sqrt{B}}{\sqrt{(A-1)(A-B)}}L_{13}-\frac{i\sqrt{A}}{\sqrt{(B-1)(A-B)}}L_{14},\nonumber \\
 L'_{34} = \frac{\sqrt{B(B-1)}}{\sqrt{A(A-1)}}L_{12}-\frac{\sqrt{B(A-B)}}{\sqrt{(A-1)}}L_{13}+i\frac{\sqrt{(B-1)(A-B)}}{\sqrt{A}}L_{23},\nonumber\\
 -iL'_{14}+iL'_{23}-L'_{13}-L'_{24} = \frac{2}{\epsilon^2}\left( \frac{i(A+B-1)}{\sqrt{AB(A-1)(B-1)}}L_{12}+\frac{i\sqrt{B}}{\sqrt{(A-1)(A-B)}}L_{13}\right.\nonumber\\
 \qquad{} - \frac{\sqrt{A}}{\sqrt{B(B-1)(A-B)}}L_{14}+\frac{\sqrt{(B-1)}}{\sqrt{A(A-B)}}L_{23}
 \left.-\frac{i\sqrt{(A-1)}}{\sqrt{B(A-B)}}L_{24}\right),\nonumber\\
 iL'_{14}+iL'_{23}-L'_{13}+L'_{24} = \frac{2i}{\epsilon^4}
 \left(-\frac{1}{\sqrt{AB(A-1)(B-1)}}(L_{12}+L_{34})\right.\nonumber\\
\left.\qquad{} +\frac{i}{\sqrt{A(B-1)(A-B)}}(L_{14}+L_{23})-\frac{1}{\sqrt{B(A-1)(A-B)}}(L_{13}-L_{24}) \right).
\end{gather*}

{\bf Limit of 2D potential}: $V_{[1,1,1,1]}
\overset{\epsilon\to0}{\Longrightarrow}V_{[4]}$,
where
\begin{gather*}
V_{[4]}=\frac{d_1}{(x'_1+ix'_2)^2}+\frac{d_2(x'_3+ix'_4)}{(x'_1+ix'_2)^3}
+d_3\left(\frac{3({x'_3}+ix'_4)^2}{(x'_1+ix'_2)^4}-2\frac{(x'_1+ix'_2)(x'_3-ix'_4)}{(x'_1+ix'_2)^4}\right)\\
\hphantom{V_{[4]}=}{}+
 d_4\frac{4(x'_1+ix'_2)({x'_1}^2+{x'_2}^2)+2(x'_3+ix'_4)^3
}{(x'_1+ix'_2)^5},\\
 a_1=-\frac{d_4}{4A^2B^2\epsilon^{12}}-\frac{d_3}{2AB^2\epsilon^{10}}-\frac{d_2}{4AB\epsilon^8}-\frac{d_1}{2AB\epsilon^6},\\
 a_2=-
\frac{d_4}{4(1-A)^2(1-B)^2\epsilon^{12}}+\frac{d_3}{2(1-A)(1-B)^2\epsilon^{10}}-\frac{d_2}{4(1-A)(1-B)\epsilon^8},\\
 a_3=-\frac{d_4}{4A^2(1-A)^2(A-B)^2\epsilon^{12}},\\
 a_4=-\frac{d_4}{4B^2(1-B)^2(A-B)^2\epsilon^{12}}-\frac{d_3}{2B^2(1-A)^2(A-B)\epsilon^{10}}.
 \end{gather*}

In these coordinates a basis for the conformal symmetry algebra is $H$, $Q_1$, $Q_2$, where
\begin{gather*} Q_1=\frac14(L_{14}+L_{23}-iL_{13}+iL_{24})^2+4a_3\left(\frac{x_1+ix_2}{x_3+ix_4}\right)+4a_4\left(\frac{x_1+ix_2}{x_3+ix_4}\right)^2,\\
 Q_2=\frac12\{L_{23}+L_{14}-iL_{13}+iL_{24},L_{12}+L_{34}\}+\frac14(L_{14}-L_{23}+iL_{13}+iL_{24})^2\\
\hphantom{Q_2=}{}
 +2a_1\left(\frac{x_1+ix_2}{x_3+ix_4}\right) +a_2\left(2\frac{x_1-ix_2}{x_3+ix_4}-\left(\frac{x_1+ix_2}{x_3+ix_4}\right)^2\right)\\
\hphantom{Q_2=}{}
+2a_3\left(6\left(\frac{x_1^2+x_2^2}{(x_3+ix_4)^2}\right)-\left(\frac{x_1+ix_2}{x_3+ix_4}\right)^3\right)\\
\hphantom{Q_2=}{}
-4a_4\left(\left(\frac{x_1-ix_2}{x_3+ix_4}\right)^2-3\left(\frac{(x_1+ix_2)^2(x_1-ix_2)}{(x_3+ix_4)^3}+\frac14
\left(\frac{x_1+ix_2}{x_3+ix_4}\right)^4\right)\right).
\end{gather*}

{\bf Basis of conformal symmetries for original system}: $\{H_0+V_{[1,1,1,1]},Q_{12},Q_{13}\}$,
where $Q_{jk}=(x_j\partial_{x_k}-x_k\partial_{x_j})^2+a_j\frac{x_k^2}{x_j^2}+a_k\frac{x_j^2}{x_k^2}$, $1\le j<k\le 4$.

{\bf Contraction of basis}: $H_0+V_{[1,1,1,1]}\to H_0'+V_{[4]}$,
\begin{gather*}\epsilon^8 Q_{12}\sim \frac{-1}{4(A-1)(B-1)AB}(L'_{13}-L'_{24}+iL'_{23}+iL'_{14})^2
+\frac{4d_3(x'_3+ix'_4)}{AB(A-1)(B-1)(x'_1+ix'_2)}\\
\hphantom{\epsilon^8 Q_{12}\sim}{}
 +\frac{d_4}{4AB(A-1)(B-1)}\left[\frac{(x'_3+ix'_4)^2}{(x'_1+ix'_2)^2}+2\frac{x_3'-ix_4'}{x_1'+ix_2'}\right],\\
 \epsilon^5\left(Q_{12}-\frac{B-A}{(1-B)A}Q_{13}\right)\sim \frac{-i}{4AB(B-1)}\\
 \hphantom{\epsilon^5\left(Q_{12}-\frac{B-A}{(1-B)A}Q_{13}\right)\sim}{}
 \times\{L'_{13}-L'_{24}+iL'_{23}+iL'_{14},
L'_{14}+iL'_{13}-L'_{23}+iL'_{14}\}\\
\hphantom{\epsilon^5\left(Q_{12}-\frac{B-A}{(1-B)A}Q_{13}\right)\sim}{}
+\frac{(A+1)d_1}{2(B-1)A^2}+\frac{d_2}{2(B-1)AB}\frac{x'_3+ix'_4}{x'_1+ix'_2}\\
\hphantom{\epsilon^5\left(Q_{12}-\frac{B-A}{(1-B)A}Q_{13}\right)\sim}{}
+\frac{d_3}{2(B-1)AB}\left[ 3\frac{(x'_3+ix'_4)^2}{(x'_1+ix'_2)^2}
-2\frac{x'_3-ix'_4}{x'_1+ix'_2}\right]\\
\hphantom{\epsilon^5\left(Q_{12}-\frac{B-A}{(1-B)A}Q_{13}\right)\sim}{}
+\frac{d_4}{(B-1)(A-1)B}\left[\frac{(x'_3+ix'_4)^3}{(x'_1+ix'_2)^3}-2\frac{{x'_3}^2+{x'_4}^2}{(x'_1+ix'_2)^2}\right].
\end{gather*}
The second limit is equivalent to the contracted Hamiltonian, not an independent basis element.

\subsection[$\protect{[2,2]}$ to $\protect{[4]}$ contraction]{$\boldsymbol{[2,2]}$ to $\boldsymbol{[4]}$ contraction}

\vspace{-5mm}

\begin{gather*}
 L'_{12} = i\left(1+\frac{2}{\epsilon}-\frac{1}{2\epsilon^2}\right)L_{12}+\frac{1}{\epsilon}\left(1-\frac{3}{4\epsilon}+\frac{1}{4\epsilon^2}\right)L_{13}
 +\frac{i}{4\epsilon^2}\left(3-\frac{1}{\epsilon}\right)L_{14} \\
 \hphantom{L'_{12} =}{}
 + \frac{i}{4\epsilon^2}\left(3-\frac{1}{\epsilon}\right)L_{23}+\left(3-\epsilon+\frac{3}{4\epsilon^2}-
 \frac{1}{4\epsilon^3}\right)L_{24}+i\left(\frac{3\epsilon}{2}-2+\frac{1}{\epsilon}-\frac{1}{2\epsilon^2}\right)L_{34}, \\
 L'_{12}+iL'_{24} = \epsilon(L_{13}-iL_{14}),\qquad L'_{13}+iL'_{34}=\epsilon(L_{23}-iL_{24}), \\
 L'_{14} = (-1+\epsilon)L_{12}+i(1-\epsilon)L_{13}+(1+\epsilon)L_{14},\qquad L'_{23}-L'_{14}=-L_{14}+L_{23}, \\
L'_{13}+L'_{24} = \left(\frac12-\frac{1}{\epsilon}\right)L_{12}+\frac{i}{\epsilon}L_{13}+\frac12 L_{14}+\frac12L_{23}+\left(2+\frac{i}{\epsilon}\right)L_{24}+\left(\epsilon-\frac12+\frac{1}{\epsilon}\right)L_{34}.
\end{gather*}

{\bf Coordinate implementation}:
 \begin{gather*}
 x_1=\frac{1}{2}\left(\frac{1}{\epsilon}+\frac{1}{\epsilon^2}\right)(x'_1-ix'_4)+\frac{\epsilon}{2}(x'_1+ix'_4)
 -\left(1+\frac{1}{2\epsilon}\right)(x'_2-ix'_3)+\frac{1}{2}(\epsilon-1)(x'_2+ix'_3),\\
 x_2=\frac{i}{2}\left(\frac{1}{\epsilon}-\frac{1}{\epsilon^2}\right)(x'_1-ix'_4)-\frac{i\epsilon}{2}(x'_1+ix'_4)
 -i\left(1-\frac{1}{2\epsilon}\right)(x'_2-ix'_3)+\frac{i}{2}(\epsilon+1)(x'_2+ix'_3),\\
 x_3=\frac{1}{2}\left(\frac{1}{\epsilon}-\frac{1}{\epsilon^2}\right)(x'_1-ix'_4)+\left(-\frac12+\frac{1}{\epsilon}\right)(x'_2-ix'_3),\\
 x_4=\frac{i}{2}\left(\frac{1}{\epsilon}+\frac{1}{\epsilon^2}\right)(x'_1-ix'_4)-i\left(\frac12+\frac{1}{\epsilon}\right)(x'_2-ix'_3).
 \end{gather*}

{\bf Limit of 2D potential}: $V_{[2,2]}
\overset{\epsilon\to0}{\Longrightarrow}V'_{[4]}$,
\begin{gather*}
V'_{[4]}=\frac{e_1}{(x'_1-ix'_4)^2}+\frac{e_2(x'_2-ix'_3)}{(x'_1-ix'_4)^3}
+e_3\left(\frac{3({x'_2}-ix'_3)^2}{(x'_1-ix'_4)^4}+2\frac{(x'_1-ix'_4)(x'_2+ix'_3)}{(x'_1-ix'_4)^4}\right)\\
\hphantom{V'_{[4]}=}{} +
 e_4\left(\frac{4(x'_1-ix'_4)({x'_2}^2+{x'_3}^2)+2(x'_2-ix'_3)^3
}{(x'_1-ix'_4)^5} \right) \qquad\!\! (\text{conformally equivalent to $V[4]$}),\\
 b_1=\frac{e_1}{\epsilon^4}+2\frac{e_4}{\epsilon^7},\qquad\!\! b_2=-\frac{e_2}{4\epsilon^6}-\frac{e_3}{2\epsilon^7}-\frac{e_4}{\epsilon^8},
 \qquad\!\!
b_3=2\frac{e_3}{\epsilon^6}-2\frac{e_4}{\epsilon^7},\qquad\!\! b_4=-\frac{e_2}{4\epsilon^6}+\frac{3e_3}{2\epsilon^7}-\frac{e_4}{\epsilon^8}.
\end{gather*}

{\bf Basis of conformal symmetries for original system}: $\{H_0+V_{[2,2]},Q_1,Q_3\}$.

{\bf Contraction of basis}: $H_0+V_{[2,2]}\to H'_0+V'_{[4]}$,
\begin{gather*}
-4\epsilon^4\left( Q_1+\frac{k_4}{\epsilon^6}-\frac{k_3}{2\epsilon^5}\right)\to (iL'_{13}-L'_{12}-iL'_{24}-L'_{34})^2\\
\qquad{}+k_2+4k_3\frac{x'_2-ix'_3}{x'_1-ix'_4}-4k_4\frac{(x'_2-ix'_3)^2}{(x'_1-ix'_4)^2},\\
 \epsilon^3 \left(Q_3-\frac{2k_4}{\epsilon^7}+\frac{k_3}{\epsilon^6}+\frac{k_1}{2\epsilon^4}\right)\to \frac{i}{2}
\{L'_{23}-L'_{14},(L'_{12}-iL'_{13}+L'_{24}+L'_{34}\}\\
\qquad{} +k_1\frac{(x'_2-ix'_3)}{(x'_1-ix'_4)}+
k_2\frac{(x'_2-ix'_3)^2}{(x'_1-ix'_4)^2}
+k_3\frac{3(x'_2-ix'_3)^3+2({x'_2}^2+{x'_3}^2)(x'_1-ix'_4)}{(x'_1-ix'_4)^3}\\
\qquad{} -2k_4(x'_2-ix'_3)\frac{(x'_2-ix'_3)^3+2({x'_2}^2+{x'_3}^2)(x'_1-ix'_4)}{(x'_1-ix'_4)^4}.
\end{gather*}
The second limit is equivalent to the contracted Hamiltonian, not an independent basis element.

 \subsection[$\protect{[3,1]}$ to $\protect{[4]}$ contraction]{$\boldsymbol{[3,1]}$ to $\boldsymbol{[4]}$ contraction}
 This contraction is not needed because the $[1,1,1,1]\to[4]$ contraction takes the $V[3,1]$ to $V[4]$.

\subsection[$\protect{[2,1,1]}$ to $\protect{[4]}$ contraction]{$\boldsymbol{[2,1,1]}$ to $\boldsymbol{[4]}$ contraction}

 This contraction is not needed because the $[1,1,1,1]\to[4]$ contraction takes $V[2,1,1]$ to $V[4]$.

\subsection[$\protect{[1,1,1,1]}$ to $\protect{[3,1]}$ contraction]{$\boldsymbol{[1,1,1,1]}$ to $\boldsymbol{[3,1]}$ contraction}

\vspace{-5mm}

\begin{gather*}
 -L'_{12}+iL'_{24} = -a\sqrt{2a^2-2}\epsilon L_{12}, \qquad
L'_{13}= -\frac{i}{\sqrt{a^2-1}}(L_{13}+aL_{12}), \\
L'_{14}+iL'_{34} = \sqrt{2}a\epsilon L_{14},\qquad
-L'_{12}+iL'_{23}=i\sqrt{2}a\epsilon L_{23},\qquad
L'_{24} = i(\sqrt{a^2-1}L_{24}-iaL_{14}),\\
-L'_{14}+iL'_{34}= \frac{\sqrt{2}}{\epsilon a\sqrt{a^2-1}}\left( L_{34}-\sqrt{a^2-1}L_{14}-iaL_{24}\right).
\end{gather*}

{\bf Coordinate implementation}:
\begin{gather*} x_1=\frac{1}{\sqrt{2}a\epsilon}(x'_1+ix'_3)+\frac{x'_2}{a}+\frac{a\epsilon}{\sqrt{2}}(x'_1-ix'_3),\qquad
x_2=\frac{i(x'_1+ix'_3)}{\sqrt{2a^2-2}\epsilon},\\
 x_3=-\frac{(x'_1+ix'_3)}{\sqrt{2a^2-2}a\epsilon}+\frac{\sqrt{a^2-1}}{a} x'_2,\qquad x_4=x'_4,\qquad a(a-1)\ne 0.
 \end{gather*}

 {\bf Limit of 2D potential}: $V_{[1,1,1,1]}
\overset{\epsilon\to0}{\Longrightarrow}V_{[31]}$,
where $V[31]$ is given by~(\ref{V[31]})
and
\begin{gather*} a_1=\frac{c_1}{2\epsilon^2}+\frac{c_3}{4a^4\epsilon^4},\qquad a_2=\frac{c_2}{4\sqrt{2}(a^2-1)^2\epsilon^3}+\frac{c_3}{4 (a^2-1)^2\epsilon^4},\\
 a_3=\frac{c_2}{4\sqrt{2}(a^2-1)^2a^2\epsilon^3}+\frac{c_3}{4 (a^2-1)^2a^4\epsilon^4},\qquad a_4=c_4.
 \end{gather*}

{\bf Basis of conformal symmetries for original system}:
$ H_0+V_{[1,1,1,1]}$, $Q_{12}$, $Q_{13}$,
where
\begin{gather*}
Q_{jk}=(x_j\partial_{x_k}-x_k\partial_{x_j})^2+a_j\frac{x_k^2}{x_j^2}+a_k\frac{x_j^2}{x_k^2},\qquad 1\le j<k\le 4.
\end{gather*}

{\bf Contracted basis}:
$H_0+V_{[1,1,1,1]}\to H_0'+V_{[3,1]}$,
\begin{gather*}
\epsilon^2\left( Q_{12}+\frac{c_3}{2a^2(a^2-1)\epsilon^4}
+\frac{\sqrt{2}c_2}{a^2(a^2-1)\epsilon^3}\right)\to -\frac{c_1}{2(a^2-1)}\\
\qquad{}
 -\frac{2c_3{x'_2}^2}{a^2(a^2-1)(x'_1+ix'_3)^2}-\frac{c_2}{2a^2(a^2-1)(x'_1+ix'_3)}-\frac{1}{2a^2(a^2-1)}(L'_{12}-iL'_{23})^2,\\
\epsilon\left(Q_{13}+a^2Q_{12}+\frac{(a^2-1)c_3}{2a^4\epsilon^4}+\frac{\sqrt{2}c_2}{8a^2\epsilon^3}+\frac{c_1(a^2-1)}{2\epsilon^2}\right)
\to \frac{\sqrt{2}c_1 x'_2}{x'_1+ix'_3}\\
\qquad{} +\frac{\sqrt{2}c_2(4{x'_2}^2+{x'_4}^2)}{4(x'_1+ix'_3)^2}+\frac{2\sqrt{2}c_3 x'_2(2{x'_2}^2+{x'_4}^2)}{(x'_1+ix'_3)^3}
+\frac{i\sqrt{2}}{2}\{L'_{13},L'_{12}-iL'_{23}\}.
\end{gather*}

\subsection[$\protect{[2,2]}$ to $\protect{[4]}$ contraction]{$\boldsymbol{[2,2]}$ to $\boldsymbol{[4]}$ contraction}

\vspace{-5mm}

\begin{gather*}
 L'_{12} = i\left(1+\frac{2}{\epsilon}-\frac{1}{2\epsilon^2}\right)L_{12}+\frac{1}{\epsilon}\left(1-\frac{3}{4\epsilon}+\frac{1}{4\epsilon^2}\right)L_{13}
 +\frac{i}{4\epsilon^2}\left(3-\frac{1}{\epsilon}\right)L_{14} \\
\hphantom{L'_{12} =}{}
+ \frac{i}{4\epsilon^2}\left(3-\frac{1}{\epsilon}\right)L_{23}+\left(3-\epsilon+\frac{3}{4\epsilon^2}-
 \frac{1}{4\epsilon^3}\right)L_{24}+i\left(\frac{3\epsilon}{2}-2+\frac{1}{\epsilon}-\frac{1}{2\epsilon^2}\right)L_{34}, \\
 L'_{12}+iL'_{24} = \epsilon(L_{13}-iL_{14}), \qquad L'_{13}+iL'_{34}=\epsilon(L_{23}-iL_{24}), \\
 L'_{14} = (-1+\epsilon)L_{12}+i(1-\epsilon)L_{13}+(1+\epsilon)L_{14},\qquad L'_{23}-L'_{14}=-L_{14}+L_{23}, \\
L'_{13}+L'_{24} = \left(\frac12-\frac{1}{\epsilon}\right)L_{12}+\frac{i}{\epsilon}L_{13}+\frac12 L_{14}+\frac12L_{23}+\left(2+\frac{i}{\epsilon}\right)L_{24}+\left(\epsilon-\frac12+\frac{1}{\epsilon}\right)L_{34}.
\end{gather*}

 {\bf Coordinate implementation}:
 \begin{gather*} x_1=\frac{1}{2}\left(\frac{1}{\epsilon}+\frac{1}{\epsilon^2}\right)(x'_1-ix'_4)+\frac{\epsilon}{2}(x'_1+ix'_4)
 -\left(1+\frac{1}{2\epsilon}\right)(x'_2-ix'_3)+\frac{1}{2}(\epsilon-1)(x'_2+ix'_3),\\
 x_2=\frac{i}{2}\left(\frac{1}{\epsilon}-\frac{1}{\epsilon^2}\right)(x'_1-ix'_4)-\frac{i\epsilon}{2}(x'_1+ix'_4)
 -i\left(1-\frac{1}{2\epsilon}\right)(x'_2-ix'_3)+\frac{i}{2}(\epsilon+1)(x'_2+ix'_3),\\
 x_3=\frac{1}{2}\left(\frac{1}{\epsilon}-\frac{1}{\epsilon^2}\right)(x'_1-ix'_4)+\left(-\frac12+\frac{1}{\epsilon}\right)(x'_2-ix'_3),\\
 x_4=\frac{i}{2}\left(\frac{1}{\epsilon}+\frac{1}{\epsilon^2}\right)(x'_1-ix'_4)-i\left(\frac12+\frac{1}{\epsilon}\right)(x'_2-ix'_3).
 \end{gather*}

{\bf Limit of 2D potential}: $V_{[2,2]}
\overset{\epsilon\to0}{\Longrightarrow}V'_{[4]}$. Conformally equivalent to $V[4]$,
 \begin{gather*}
 V'_{[4]}=\frac{e_1}{(x'_1-ix'_4)^2}+\frac{e_2(x'_2-ix'_3)}{(x'_1-ix'_4)^3}
+e_3\left(\frac{3({x'_2}-ix'_3)^2}{(x'_1-ix'_4)^4}+2\frac{(x'_1-ix'_4)(x'_2+ix'_3)}{(x'_1-ix'_4)^4}\right)\\
\hphantom{V'_{[4]}=}{}+
 e_4\left(\frac{4(x'_1-ix'_4)({x'_2}^2+{x'_3}^2)+2(x'_2-ix'_3)^3
}{(x'_1-ix'_4)^5} \right) ,\\
 b_1=\frac{e_1}{\epsilon^4}+2\frac{e_4}{\epsilon^7},\qquad\!\! b_2=-\frac{e_2}{4\epsilon^6}-\frac{e_3}{2\epsilon^7}-\frac{e_4}{\epsilon^8},
 \qquad \!\!
b_3=2\frac{e_3}{\epsilon^6}-2\frac{e_4}{\epsilon^7},\qquad\!\! b_4=-\frac{e_2}{4\epsilon^6}+\frac{3e_3}{2\epsilon^7}-\frac{e_4}{\epsilon^8}.
\end{gather*}

{\bf Basis of conformal symmetries for original system}: $\{H_0+V_{[2,2]},Q_1,Q_3\}$.

{\bf Contraction of basis}: $H_0+V_{[2,2]}\to H'_0+V'_{[4]}$,
\begin{gather*}
-4\epsilon^4\left( Q_1+\frac{k_4}{\epsilon^6}-\frac{k_3}{2\epsilon^5}\right)\to (iL'_{13}-L'_{12}-iL'_{24}-L'_{34})^2\\
\qquad{} +k_2+4k_3\frac{x'_2-ix'_3}{x'_1-ix'_4}-4k_4\frac{(x'_2-ix'_3)^2}{(x'_1-ix'_4)^2},\\
 \epsilon^3 \left(Q_3-\frac{2k_4}{\epsilon^7}+\frac{k_3}{\epsilon^6}+\frac{k_1}{2\epsilon^4}\right)\to \frac{i}{2}
\{L'_{23}-L'_{14},(L'_{12}-iL'_{13}+L'_{24}+L'_{34}\}\\
\qquad {}+k_1\frac{(x'_2-ix'_3)}{(x'_1-ix'_4)}+
k_2\frac{(x'_2-ix'_3)^2}{(x'_1-ix'_4)^2} +k_3\frac{3(x'_2-ix'_3)^3+2({x'_2}^2+{x'_3}^2)(x'_1-ix'_4)}{(x'_1-ix'_4)^3}\\
\qquad {}-2k_4(x'_2-ix'_3)\frac{(x'_2-ix'_3)^3+2({x'_2}^2+{x'_3}^2)(x'_1-ix'_4)}{(x'_1-ix'_4)^4}.
\end{gather*}
The second limit is equivalent to the contracted Hamiltonian, not an independent basis element.

\subsection{Summary of B\^ocher contractions of Laplace systems}\label{4}

\looseness=-1
This is a summary of the results of applying each
of the B\^ocher contractions to each of the Laplace conformally superintegrable systems. In many cases a single contraction gives rise to
rise to more than one result, due to the fact that the indices of the image potential can be permuted and image potential may not be
permutation invariant. The details can be found in~\cite{ArXiv2016}.
 \begin{enumerate}\itemsep=0pt
\item $[1,1,1,1]\to [2,1,1]$ contraction:
 \begin{gather*}
 V_{[1,1,1,1]}\downarrow V_{[2,1,1]}; \quad V_{[2,1,1]}\downarrow V_{[2,1,1]},V_{[2,2]},V_{[3,1]};\quad
V_{[2,2]}\downarrow V_{[2,2]},V_{[0]};
\\
V_{[3,1]}\downarrow V_{(1)},V_{[3,1]};\quad V_{[4]}\downarrow V_{[0]},V_{(2)}; \quad V_{[0]}\downarrow V_{[0]}; \quad V_{(1)}\downarrow V_{(1)},V_{(2)};
 \quad V_{(2)}\downarrow V_{(2)}.
 \end{gather*}
\item $[1,1,1,1]\to [2,2]$ contraction:
\begin{gather*}
V_{[1,1,1,1]}\downarrow V_{[2,2]}; \quad V_{[2,1,1]}\downarrow V_{[2,2]} \ (\text{special case of}\ E15);\quad
V_{[2,2]}\downarrow V_{[2,2]},V_{[0]};\\
V_{[3,1]}\downarrow V_{(1)} \ (\text{special case of}~E_{15});\quad
 V_{[4]}\downarrow V_{(2)}; \quad V_{[0]}\downarrow V_{[0]}; \\
 V_{(1)}\downarrow V_{(1)} \ (\text{special case of}~E15); \quad V_{(2)}\downarrow V_{(2)}.
 \end{gather*}
\item $[2,1,1]\to [3,1]$ contraction:
\begin{gather*} V_{[1,1,1,1]}\downarrow V_{[3,1]};\quad V_{[2,1,1]}\downarrow V_{[3,1]},V_{[0]};\quad
 V_{[2,2]}\downarrow V_{[0]};\quad
 V_{[3,1]}\downarrow V_{[3,1]},V_{[0]};\quad
 V_{[4]}\downarrow V_{[0]};\\
 V_{[0]}\downarrow V_{[0}];\quad V_{(1)}\downarrow V_{(2)};\quad V_{(2)}\downarrow V_{(2)}.
 \end{gather*}
\item $[1,1,1,1]\to [4]$ contraction:
\begin{gather*} V_{[1,1,1,1]}\downarrow V_{[4]};\quad V_{[2,1,1]}\downarrow V_{[4]};\quad
 V_{[2,2]}\downarrow V_{[0]};\quad
 V_{[3,1]}\downarrow V_{[4]};\quad
 V_{[4]}\downarrow V_{[0]},V_{[4]};\\
V_{[0]}\downarrow V_{[0]};\quad V_{(1)}\downarrow V_{(2)};\quad V_{(2)}\downarrow V_{(2)}.
\end{gather*}
\item $[2,2]\to [4]$ contraction:
 \begin{gather*} V_{[1,1,1,1]}\downarrow V_{[4]};\quad V_{[2,1,1]}\downarrow V_{[4]},V_{(2)};\quad
 V_{[2,2]}\downarrow V_{[4]},V_{[0]};\quad
 V_{[3,1]}\downarrow V_{(2)};\quad
 V_{[4]}\downarrow V_{(2)};\\
 V_{[0]}\downarrow V_{[0]},V_{(2)};\quad V_{(1)}\downarrow V_{(2)};\quad V_{(2)}\downarrow V_{(2)}.
 \end{gather*}

\item $[1,1,1,1]\to [3,1]$ contraction:
\begin{gather*} V_{[1,1,1,1]}\downarrow V_{[3,1]};\quad V_{[2,1,1]}\downarrow V_{[3,1]},V_{[0]};\quad
 V_{[2,2]}\downarrow V_{[0]};\quad
 V_{[3,1]}\downarrow V_{[3,1]},V_{[0]};\\
 V_{[4]}\downarrow V_{[0]}; \quad V_{[0]}\downarrow V_{[0]};\quad
 V_{(1)}\downarrow V_{(2)}; \quad V_{(2)}\downarrow V_{(2)}.
 \end{gather*}
\end{enumerate}

\subsection{Conformal St\"ackel transforms of the Laplace systems} \label{3.1.1}

We give the details of the description of
the Helmholtz systems that follow from the Laplace system $[1,1,1,1]$ by conformal St\"ackel transform
\begin{gather*}
V_{[1,1,1,1]}=\frac{a_1}{x_1^2}+\frac{a_2}{x_2^2}+\frac{a_3}{x_3^2}+\frac{a_4}{x_4^2}.
\end{gather*}
 We write the parameters $a_j$ def\/ining the potential $V_{[1,1,1,1]}$ as a vector: $(a_1,a_2,a_3,a_4)$. A~St\"ackel transform is generated by
 the potential
 \begin{gather*}
 U=\frac{b_1}{x_1^2}+\frac{b_2}{x_2^2}+\frac{b_3}{x_3^2}+\frac{b_4}{x_4^2}
 \end{gather*} corresponding to the vector $(b_1,b_2,b_3,b_4)$.
\begin{enumerate}\itemsep=0pt
\item The potentials $(1,0,0,0)$, and any permutation of the indices $b_j$ generate conformal St\"ackel transforms to~$S9$.
\item The potentials $(1,1,0,0)$ and $(0,0,1,1)$ generate conformal St\"ackel transforms to~$S7$.
\item The potentials $(1,1,1,1)$, $(0,1,0,1)$, $(1,0,1,0)$, $(0,1,1,0)$ and $(1,0,0,1)$ generate conformal St\"ackel transforms to $S8$.
\item The potentials $(b_1,b_2,0,0)$, $b_1b_2\ne 0$, $b_1\ne b_2$, and any permutation of the indices~$b_j$
generate conformal St\"ackel transforms to $D4B$.
\item The potentials $(1,1,a,a)$, $a\ne 0,1$, and any permutation of the indices~$b_j$.
generate conformal St\"ackel transforms to~$D4C$.
\item Each potential not proportional to one of these must generate a conformal St\"ackel transform to a~superintegrable system on a Koenigs space in the
family $K[1,1,1,1]$.
 \end{enumerate}
Similar details for all of the other Laplace systems are given in~\cite{ArXiv2016}. Here, we simply list the Helmholtz systems in each equivalence class.

\subsection{Summary of St\"ackel equivalence classes of Helmholtz systems}\label{Stackelequivclasses}
\begin{enumerate}\itemsep=0pt
\item $[1,1,1,1]$: $S9$, $S8$, $S7$, $D4B$, $D4C$, $K[1,1,1,1]$.
\item $[2,1,1]$: $S4$, $S2$, $E1$, $E16$, $D4A$, $D3B$, $D2B$, $D2C$, $K[2,1,1]$.
\item $[2,2]$: $E8$, $E17$, $E7$, $E19$, $D3C$, $D3D$, $K[2,2]$.
\item $[3,1]$: $S1$, $E2$, $D1B$, $D2A$, $K[3,1]$.
\item $[4]$: $E10$, $E9$, $D1A$, $K[4]$.
\item $[0]$: $E20$, $E11$, $E3'$, $D1C$, $D3A$, $K[0]$.
\item $(1)$: special cases of $E15$.
\item $(2)$: special cases of $E15$.
\end{enumerate}

\section{Helmholtz contractions from B\^ocher contractions}\label{3}

 We describe how B\^ocher contractions of
conformal superintegrable systems induce contractions
of Helmholtz superintegrable systems. The basic idea here is that the procedure of taking a conformal St\"ackel transform of a~conformal superintegrable system, followed
by a Helmholtz contraction yields the same result as taking a B\^ocher contraction followed by an ordinary St\"ackel transform: The
diagrams commute~\cite{KMS2016}. To describe this process we recall that each of the B\^ocher systems classif\/ied above can be
considered as an equivalence class of Helmholtz superintegrable systems under the St\"ackel transform. We now determine the
Helmholtz systems in each equivalence class and how they are related.

\begin{figure}[h] \label{figure1}
\centering
\includegraphics[scale=0.137]{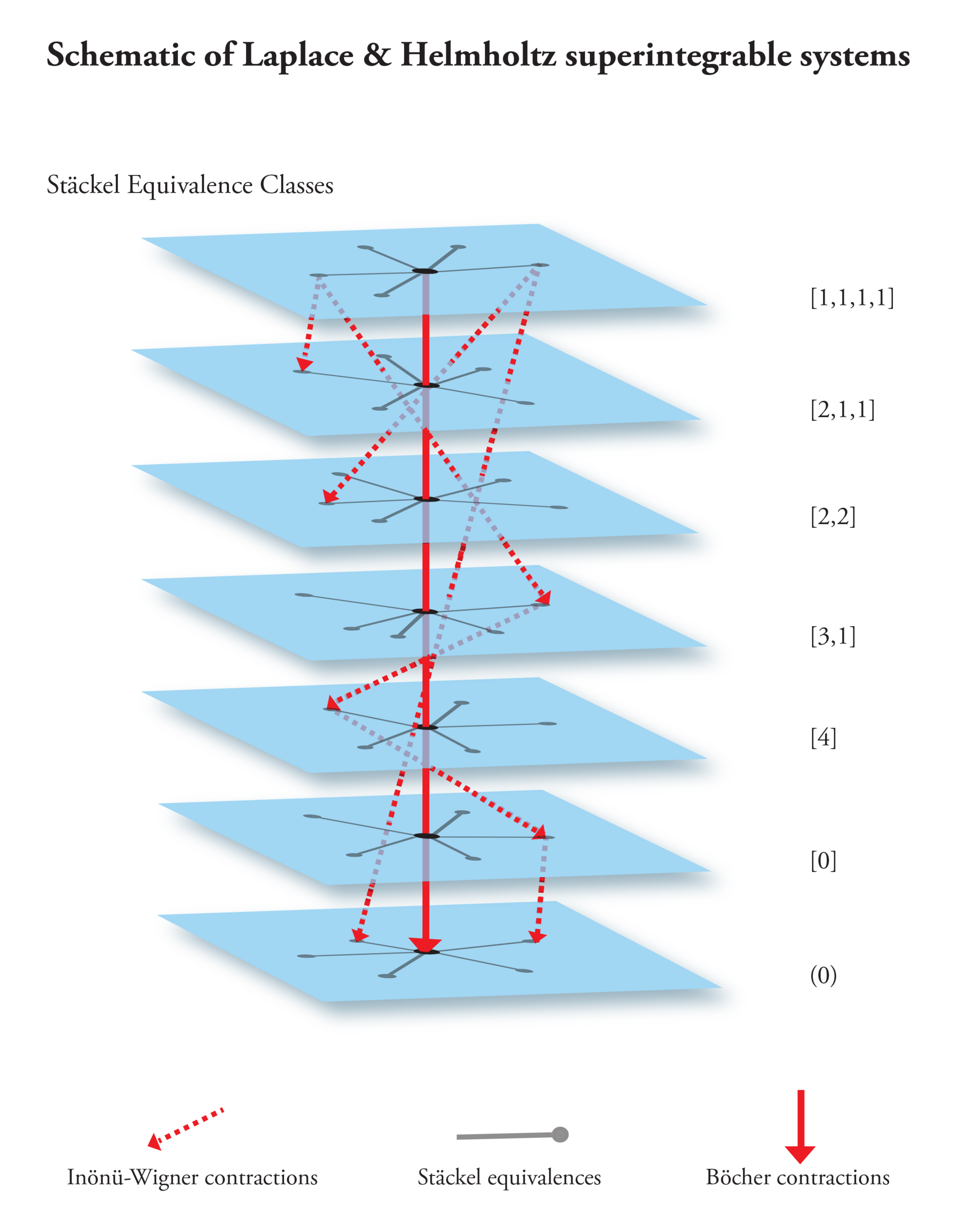}
\caption{The bigger picture.}\label{Fig1}
\end{figure}

\looseness=-1
Consider the conformal St\"ackel transforms of the conformal system $[1,1,1,1]$ with poten\-tial~$V_{[1,1,1,1]}$.
The various possibilities are listed in Section~\ref{3.1.1}. Let $H$ be the initial Hamiltonian. In
terms of tetraspherical coordinates the conformal St\"ackel transformed potential will
take the form
\begin{gather*} V=\frac{\frac{a_1}{x_1^2}+\frac{a_2}{x_2^2}+\frac{a_3}{x_3^2}+\frac{a_4}{x_4^2}}{\frac{A_1}{x_1^2}+\frac{A_2}{x_2^2}+\frac{A_3}{x_3^2}+\frac{A_4}{x_4^2}}
=\frac{V_{[1,1,1,1]}}{F({\bf x},{\bf A})},\qquad
 F({\bf x},{\bf A})=\frac{A_1}{x_1^2}+\frac{A_2}{x_2^2}+\frac{A_3}{x_3^2}+\frac{A_4}{x_4^2},
 \end{gather*}
and the transformed Hamiltonian will be
${\hat H}=\frac{1}{F({\bf x},{\bf A})}H$,
where the transform is determined by the f\/ixed vector $(A_1,A_2,A_3,A_4)$.
Now we apply the B\^ocher contraction $[1,1,1,1]\to [2,1,1]$ to this system.
In the limit as $\epsilon\to 0$
the potential $V_{[1,1,1,1]}\to V_{[2,1,1]}$, (\ref{V[211]}), and $H\to H'$ of the $[2,1,1]$ system. Now consider
$F({\bf x}(\epsilon),{\bf A})= V'({\bf x}',A)\epsilon^\alpha+O(\epsilon^{\alpha+1})$,
where the integer exponent $\alpha$ depends upon our choice of~$\bf A$. We will provide the theory to show that the system def\/ined by Hamiltonian
$ {\hat H}'=\lim\limits_{\epsilon\to 0}\epsilon^\alpha {\hat H}(\epsilon)=\frac{1}{V'({\bf x}',A)}H'$
is a superintegrable system that arises from the system $[2,1,1]$ by a conformal St\"ackel transform induced by the potential $V'({\bf x}',A)$.
Thus the Helmholtz superintegrable system with potential $V=V_{1,1,1,1}/F$ contracts to the Helmholtz superintegrable system with potential $V_{[2,1,1]}/V'$.
The contraction is induced by a~generalized In\"on\"u--Wigner Lie algebra contraction of the conformal algebra $\mathfrak{so}(4,\C)$.
In this case the possibilities for~$V'$ can be computed easily from the limit expressions~(\ref{coordlimit}).
Then the~$V'$ can be identif\/ied with a~$[2,1,1]$ potential from the list in Section~\ref{3.2}. The results follow. For each~$\bf A$
corresponding to a constant curvature or Darboux superintegrable system~$O$ we list the contracted system~$O'$ and~$\alpha$. For Koenigs spaces we will not go into detail
but merely give the contraction for a ``generic'' Koenigs system: One for which there are no rational numbers~$r_j$, not all~$0$, such that
$\sum\limits_{j=1}^4r_jA_j=0$. This ensures that the contraction is also ``generic''. The schematic to keep in mind
that relates conformal and regular St\"ackel transforms, B\^ocher contractions, Helmholtz and Laplace superintegrable
systems is Fig.~\ref{Fig1}.

\begin{Example}
 In Section \ref{3.1.1}, consider St\"ackel transform $(1,0,0,0)$, i.e., $U=1/x_1^2$.
The transformed system is
\begin{gather*}
H=\frac{1}{\frac{1}{x_1^2}}\left(\sum_{i=1}^4 \partial_{x_i}^2\right)+ \frac{1}{\frac{1}{x_1^2}}\left(\frac{a_1}{x_1^2}+\frac{a_2}{x_2^2}+\frac{a_3}{x_3^2}+\frac{a_4}{x_4^2}\right),
\end{gather*}
which is $S9$. Now take the $[1,1,1,1]\to [2,1,1]$ B\^ocher contraction, equation~(\ref{V[211]}). The sum of the derivatives in $H$ goes to $\sum\limits_{i=1}^4 \partial_{x'_i}^2$ and the numerator of the potential goes to
equation~(\ref{V[211]}). However, the denominator $1/x_1^2$ goes as
$1/x_1^2=-2 \epsilon^2/((x_1'+ix_2')^2 +O(\epsilon^6)$, so $\alpha=2$.
 Thus, if we set
$H'=\epsilon^2 H$ and go to the limit as $\epsilon \to 0$, we get
a contracted system with potential
$b_1+b_2(x^2+y^2)+b_3/x^2+b_4/y^2$
in Cartesian coordinates, up to a~scalar factor $-2$. This is~$E1$.
\end{Example}

The complicated details of the possible Helmholtz contractions induced by B\^ocher contractions of Laplace
systems are presented in~\cite{ArXiv2016}. Here, we summarize the results. In many cases a single contraction gives rise
to more than one result, due to the fact that the indices of the image potential can be permuted and image potential may not be
permutation invariant.

\subsection{Summary of Helmholtz contractions} \label{Helmholtzcontractions}

The superscript for each targeted Helmholtz system is the value of~$\alpha$. In each table, correspon\-ding to a~single Laplace equation equivalence class, the top line is
a list of the Helmholtz systems in the class, and the lower lines are the target systems under the B\^ocher contraction.

\begin{table}[th!]\centering
\caption{$[1,1,1,1]$ equivalence class contractions.}\label{table1}\vspace{1mm}

\begin{tabular}{ccccccc}
\hline
{\rm contraction} & ${S_9}$ & $S_7$ & $S_8$ & $D_4B$ & $D_4C$ & $K[1111]$\\
\hline
$[1111]\downarrow[211]$ & $E_1^2$ & $S_4^0$ & $S_4^0$ & $E_1^2$ & $S_4^0$& $D_4A^0$\tsep{2pt}\\
&$S_2^0$ & $S_2^0$ & $E_{16}^0$ & $D_4A^0$ & $D_4A^0$\\
&&&&$S_2^0$&\\
\hline
$[1111]\downarrow[22]$& $E_7^2$ & $E_{19}^4$ & $E_{17}^4$ & $E_7^2$ & $E_{19}^1$ & $E_7^2$\tsep{2pt}\\
&&$E_7^2$& $E_{19}^4$\\
& &$E_{17}^2$ &\\
\hline
$[1111]\downarrow[31]$ & $E_2^2$ & $S_1^0$& $S_1^0$& $S_1^0$ & $S_1^0$& $S_1^0$\tsep{2pt}\\
&$S_1^0$& $E_2^2$& $E_2^2$& $E_2^2$\\
\hline
$[1111]\downarrow[4]$ & $E_{10}^6$ & $E_{10}^6$ & $E_{10}^6$ & $E_{10}^6$ & $E_{10}^6$& $E_{10}^6$\tsep{2pt}\\
\hline
$[22]\downarrow[4]$ & $E_{10}^4$ & $E_9^6$ & $E_{10}^5$ & $E_{10}^4$ & $E_{10}^5$ & $E_{10}^4$\tsep{2pt}\\
&&$E_{10}^4$ &  \\
&&$E_9^5$&\\
\hline
$[211]\downarrow[31]$ & $E_2^6$ & $S_1^0$ & $S_1^0$ & $S_1^0$ & $S_1^0$ & $S_1^0$ \tsep{2pt}\\
&$E_2^4$ & $E_2^4$ & $E_2^8$ & $E_2^6$\\
&$S_1^0$ &&& $E_2^4$\\
\hline
\end{tabular}
\end{table}

\begin{table}[th!]\centering
\caption{$[2,1,1]$ equivalence class contractions.}\label{table2}\vspace{1mm}
\begin{tabular}{cccccccccc}
\hline
{\rm contraction} & ${S_4}$ & $S_2$ & $E_1$ & $E_{16}$ & $D_4A$ & $D_3B$ & $D_2B$ & $D_2C$ & $K[211]$\\
\hline
$[1111]\downarrow[211]$ & $S_4^0$ & $S_2^0$ & $E_1^2$ & $E_{16}^4$ & $D_4A^0$ & $E_1^2$ & $S_2^0$ & $S_4^0$ & $S_4^0$\tsep{2pt}\\
&$E_{17}^4$ & $E_8^2$ & $E_8^0$ & $E_{17}^0$ & $E_8^2$ & $D_3C^0$ & $E_8^0$ & $E_{17}^0$ & $D_3C^0$\\
&$S_1^0$ & $S_1^0$ & $E_2^2$ & $E_2^2$ & $S_1^0$ & $E_2^2$ & $S_1^0$ & $S_1^0$ & $S_1^0$\\
&& $E_2^2$ &&&& $D_1B^3$ & $E_2^2$ &&\\
\hline
$[1111]\downarrow[22]$ & $E_{17}^4$ & $E_{8}^2$ & $E_{8}^2$ & $E_{17}^4$ & $E_{7}^2$ & $E_8^2$ & $E_7^2$ & $E_{19}^4$ & $E_7^2$\tsep{2pt}\\
&&&&&$E_8^2$ & $E_{17}^2$ & $E_8^2$ & $E_{17}^4$ &\\
\hline
$[1111]\downarrow[31]$ & $S_1^0$ & $S_1^0$ & $E_2^2$ & $E_2^2$ & $S_1^0$ & $E_2^2$ & $E_1^2$ & $S_1^0$ & $S_1^0$\tsep{2pt}\\
&&&&&&$D_1B^3$&&&\\
&${E_3'}^2$ & ${E_3'}^2$ & ${E_3'}^2$ & ${E_3'}^2$ & ${E_3'}^2$ & ${E_3'}^2$ & ${E_3'}^2$ & ${E_3'}^2$ & ${E_3'}^2$\\
&&&&&$D_1C^3$ & $D_1C^3$ & $D_1C^3$ &&\\
\hline
$[1111]\downarrow[4]$ & $E_{10}^6$ & $E_{10}^6$ & $E_{10}^6$ & $E_{10}^6$ & $E_{10}^6$ & $E_{10}^6$ & $E_{10}^6$ & $E_{10}^6$ & $E_{10}^6$\tsep{2pt}\\
&&&&& $E_9^8$ & $E_9^8$ & $E_9^8$ & $E_9^8$ &\\
\hline
$[22]\downarrow[4]$ & $E_{10}^5$ & $E_{10}^4$ & $E_{10}^4$ & $E_{10}^5$ & $E_{10}^4$ & $E_{10}^4$ & $E_{10}^4$ & $E_{10}^4$ & $E_{10}^4$\tsep{2pt}\\
&&&&&&$E_{10}^5$ & $E_{10}^5$ &&\\
& \multicolumn{9}{c}{St\"{a}ckel transforms of $V(2)$}\\
\hline
$[211]\downarrow[31]$ & $S_1^0$ & $S_1^0$ & $E_2^6$ & $E_2^8$ & $S_1^0$ & $E_2^6$ & $S_1^0$ & $S_1^0$ & $S_1^0$\tsep{2pt}\\
&&$E_2^5$&&&&&$E_2^5$&&\\
& ${E_3'}^8$ & ${E_3'}^6$ & ${E_3'}^4$ & ${E_3'}^4$ & ${E_3'}^6$ & ${E_3'}^6$ & ${E_3'}^4$ & ${E_3'}^4$ & ${E_3'}^4$\\
\hline
\end{tabular}
\end{table}

\begin{table}[th!]\centering
\caption{$[2,2]$ equivalence class contractions.}\label{table3}\vspace{1mm}
\begin{tabular}{cccccccc}
\hline
{\rm contraction} & $E_8$ & $E_{17}$ & $E_7$ & $E_{19}$ & $D_3C$ & $D_3D$ & $K[22]$\\
\hline
$[1111]\downarrow[211]$ & $E_8^0$ & $E_{17}^0$ & $E_7^0$ & $E_{19}^0$ & $D_3C^0$ & $E_7^2$ & $D_3C^0$\tsep{2pt}\\
&${E_3'}^2$ & ${E_3'}^2$ & ${E_3'}^2$ & ${E_3'}^2$ & ${E_3'}^2$ & ${E_3'}^2$ & ${E_3'}^2$\\
\hline
$[1111]\downarrow[22]$ & $E_{8}^2$ & $E_{17}^4$ & $E_{7}^2$ & $E_{19}^4$ & $E_{8}^2$ & $E_8^2$ & $E_7^2$\tsep{2pt}\\
& ${E_3'}^2$ & $E_{11}^2$ & ${E_3'}^2$ & $E_{11}^2$ & $E_{11}^2$ & $E_{11}^2$ & $E_{11}^2$\\
\hline
$[1111]\downarrow[31]$ & ${E_3'}^2$ & ${E_3'}^2$ & ${E_3'}^2$ & ${E_3'}^2$ & ${E_3'}^2$ & ${E_3'}^2$ & ${E_3'}^2$ \tsep{2pt}\\
&&&& $E_{11}^4$ & $D_1C^3$ & $D_1C^3$ &\\
&&&& $E_{20}^4$ &&&\\
\hline
$[1111]\downarrow[4]$ & ${E'}_{3}^6$ & ${E'}_{3}^6$ & ${E'}_{3}^6$ & ${E'}_{3}^6$ & ${E'}_{3}^6$ & ${E'}_{3}^6$ & ${E'}_{3}^6$\tsep{2pt}\\
&&&$E_{11}^8$ & $E_{11}^8$ & $E_{11}^8$ & $E_{11}^8$&\\
\hline
$[22]\downarrow[4]$ & $E_{10}^4$ & $E_{10}^5$ & $E_{10}^4$ & $E_{10}^5$ & $E_{10}^4$ & $E_{10}^4$ & $E_{10}^4$\tsep{2pt}\\
&&&$E_9^5$& $E_9^6$&&&\\
&${E_3'}^2$ & $E_{11}^1$ & ${E_3'}^2$ & $E_{11}^1$ & $E_{11}^1$ & $E_{11}^1$ & $E_{11}^1$\\
&&& $E_{11}^3$ & $E_{20}^4$ &&&\\
\hline
$[211]\downarrow[31]$ & ${E'_3}^4$ & ${E'}_3^4$ & ${E'_3}^2$ & ${E_3'}^2$ & ${E'_3}^4$ & $D_1C^2$ & $D_1C^2$\tsep{2pt}\\
& ${E_3'}^6$ & ${E_3'}^6$ & ${E_3'}^6$ & ${E_{20}}^4$ & ${E_3'}^6$ & ${E_3'}^6$ & ${E_3'}^6$\\
&&&&&$D_1C^9$&&\\
\hline
\end{tabular}
\end{table}

\begin{table}[th!]\centering
\caption{$[3,1]$ equivalence class contractions.}\label{table4}\vspace{1mm}
\begin{tabular}{cccccc}
\hline
{\rm contraction} & $S_1$ & $E_{2}$ & $D_1B$ & $D_2A$ & $K[31]$\\
\hline
$[1111]\downarrow[211]$ &\multicolumn{5}{c}{St\"{a}ckel transforms of $V(1)$}\\
&$S_1^0$ & $E_2^2$ & $E_2^2$ & $E_2^2$ & $S_1^0$\\
&&& $D_1B^3$ & $D_2A^4$ &\\
\hline
$[1111]\downarrow[22]$ & \multicolumn{5}{c}{St\"{a}ckel transforms of $V(1)$}\\
\hline
$[1111]\downarrow[31]$ & $S_1^0$ & $E_2^2$ & $E_2^2$ & $E_2^2$& $S_1^0$\tsep{2pt}\\
&&& $D_1B^3$ &&\\
&${E_3'}^2$ & ${E_3'}^2$ & ${E_3'}^2$ & ${E_3'}^2$ & ${E_3'}^2$ \\
&&&$D_1C^3$&&\\
\hline
$[1111]\downarrow[4]$ & ${E}_{10}^6$ & ${E}_{10}^6$ & ${E}_{10}^6$ & ${E}_{10}^6$ & ${E}_{10}^6$\tsep{2pt}\\
&&&$E_9^8$&&\\
\hline
$[22]\downarrow[4]$ & \multicolumn{5}{c}{St\"{a}ckel transforms of $V(2)$}\\
\hline
$[211]\downarrow[31]$ & $S_1^0$ & $E_2^6$ & $E_2^6$ & $E_2^6$ & $S_1^0$\tsep{2pt}\\
&&$E_2^2$ & $S_1^1$ & $S_1^0$ &\\
&${E_3'}^4$ & ${E_3'}^6$ & ${E_3'}^6$ & ${E_3'}^6$ & ${E_3'}^4$\\
\hline
\end{tabular}
\end{table}

\begin{table}[th!]\centering
\caption{$[4]$ equivalence class contractions.}\label{table5}\vspace{1mm}
\begin{tabular}{ccccc}
\hline
{\rm contraction} & $E_{10}$ & $E_{9}$& $D_1A$& $K[4]$\\
\hline
$[1111]\downarrow[211]$ & ${E_3'}^2$ & $E_{11}^2$ & $E_{20}^4$ & ${E_3'}^2$\tsep{2pt}\\
&& ${E_3'}^2$& ${E_3'}^2$ &\\
&\multicolumn{4}{c}{St\"{a}ckel transforms of $V(2)$}\\
\hline
$[1111]\downarrow[22]$ & \multicolumn{4}{c}{St\"{a}ckel transforms of $V(2)$}\\
&${E_3'}^2$ & ${E_3'}^2$ & $D_1C^2$ & $D_3A^2$\\
\hline
$[1111]\downarrow[31]$ & ${E_3'}^2$ & ${E_3'}^2$ & ${E_3'}^2$ & ${E_3'}^2$\tsep{2pt}\\
&$E_{11}^2$ &&&\\
\hline
$[1111]\downarrow[4]$ & ${E_3'}^6$ & ${E_3'}^6$ & ${E_3'}^6$ & ${E_3'}^6$\tsep{2pt}\\
&$E_{11}^8$&&&\\
&$E_{10}^6$ & $E_{10}^6$ & $E_{10}^6$ & $E_{10}^6$\\
&$E_9^8$&&&\\
\hline
$[22]\downarrow[4]$& \multicolumn{4}{c}{St\"{a}ckel transforms of $V(2)$}\\
\hline
$[211]\downarrow[31]$ & ${E_3'}^1$ & ${E_3'}^1$ & ${E_3'}^{-1}$ & ${E_3'}^{-1}$\tsep{2pt}\\
&${E_3'}^4$ & ${E_3'}^5$ & ${E_3'}^4$ & ${E_3'}^3$\\
&${E_3'}^6$ & ${E_3'}^6$ & ${E_3'}^6$ & ${E_3'}^6$\\
\hline
\end{tabular}
\end{table}

\begin{table}[th!]\centering
\caption{$[0]$ equivalence class contractions.}\label{table6}\vspace{1mm}
\begin{tabular}{ccccccc}
\hline
{\rm contraction} & $E_{20}$ & $E_{11}$ & $E_3'$ & $D_1C$ & $D_3A$ & $K[0]$\\
\hline
$[1111]\downarrow[211]$ & ${E_3'}^2$ & ${E_{3}'}^2$ & ${E_3'}^2$ & ${E_3'}^2$ & ${E_3'}^2$ & ${E_3'}^2$\tsep{2pt}\\
&$E_{11}^3$& $E_{11}^3$ && $D_1C^3$ & $D_1C^3$&\\
\hline
$[1111]\downarrow[22]$ & $E_{11}^2$ & $E_{11}^2$ & ${E_3'}^2$ & $E_{11}^2$ & $E_{11}^2$ & $E_{11}^2$\tsep{2pt}\\
&&&&&${E_3'}^2$ & ${E_3'}^2$\\
\hline
$[1111]\downarrow[31]$ & ${E_3'}^2$ & ${E_3'}^2$ & ${E_3'}^2$ & ${E_3'}^2$ & ${E_3'}^2$ & ${E_3'}^2$\tsep{2pt}\\
&&&&$D_1C^3$& $D_1C^3$&\\
\hline
$[1111]\downarrow[4]$ & ${E_3'}^6$ & ${E_3'}^6$ & ${E_3'}^6$ & ${E_3'}^6$ & ${E_3'}^6$ & ${E_3'}^6$\tsep{2pt}\\
&$E_{11}^8$ & $E_{11}^8$ && $E_{11}^8$ & $E_{11}^8$ &\\
\hline
$[22]\downarrow[4]$ & ${E_3'}^4$ & ${E_3'}^4$ & ${E_3'}^4$ & ${E_3'}^4$ & ${E_3'}^4$ & ${E_3'}^4$ \tsep{2pt}\\
&$E_{11}^5$ & $E_{11}^5$&& $E_{11}^5$&\\
\hline
$[211]\downarrow[31]$ & ${E_3'}^6$ & ${E_3'}^6$ & ${E_3'}^6$ & ${E_3'}^6$ & ${E_3'}^6$ & ${E_3'}^6$\tsep{2pt}\\
&&&& $D_1C^9$ &&\\
\hline
\end{tabular}
\end{table}

\section{Conclusions and discussion}
The use of Lie algebra contractions based on the symmetry groups of constant
curvature spaces to construct quadratic algebra contractions of 2nd order 2D Helmholtz superintegrable systems
is esthetically pleasing but incomplete, because it doesn't satisfactorily account for Darboux and Koenigs spaces.
Also the hierarchy of contractions is confusing. The situation is clarif\/ied when one extends these systems to
2nd order Laplace
conformally superintegrable systems with conformal symmetry algebra. Classes of St\"ackel equivalent Helmholtz
superintegrable systems are now recognized as corresponding to a single Laplace superintegrable system on f\/lat space
with underlying conformal symmetry algebra $\mathfrak{so}(4,\C)$.
The conformal Lie algebra contractions are induced by B\^ocher limits of $\mathfrak{so}(4,\C)$ to itself
associated with invariants of quadratic forms.
Except for one special class they generalize all of the Helmholtz contractions derived earlier. In particular, contractions of Darboux and
Koenigs systems can be described easily. All of the concepts introduced in this paper are clearly also applicable for
dimensions $n\ge 3$, see~\cite{CKP2015}. The conceptual picture is Fig.~\ref{Fig1}.

The special class that is missing in the present paper is the class of contractions to systems with degenerate Hamiltonians, i.e., systems for which the determinant of the metric tensor
is zero. In a paper under preparation we will show that these limits correspond to contractions of
$\mathfrak{so}(4,\C)$ to $\mathfrak{e}(3,\C)$ and lead to time-dependent conformally superintegrable systems (Schr\"odinger equations) with potential. We will examine the relations between the
contractions in classif\/ied in~\cite{HKMS, KM2014} and show that they are properly contained in those induced by $\mathfrak{so}(4,\C)$.
From Theo\-rem~\ref{theorem1} we know that the potentials of all Helmholtz superintegrable systems are completely determined by their
free quadratic algebras, i.e., the symmetry algebra that remains when the parameters in the potential
are set equal to~0. Thus for classif\/ication purposes it is enough to classify free abstract quadratic algebras.
We will give a classif\/ication of abstract free nondegenerate quadratic algebras and their abstract
contractions and discuss which of these abstract systems and contractions correspond to physical systems.

In papers under preparation we will 1) give a precise def\/inition of B\"ocher contractions and introduce other
methods of constructing them,
2)~apply the B\^ocher construction to degenerate (1-parameter) Helmholtz
superintegrable systems (which admit a 1st order symmetry),
3)~give a complete classif\/ication of free abstract degenerate quadratic algebras and identify which
of those correspond to free 2nd order superintegrable systems,
4)~classify abstract contractions of degenerate quadratic algebras and identify which of those
correspond to geometric contractions of Helmholtz superintegrable systems.

\looseness=-1
We note that by taking contractions step-by-step from a model of the $S_9$ quadratic algebra we can recover the Askey scheme~\cite{KMP2014}.
However, the contraction method is more general. It applies to all special functions
that arise from the quantum systems via separation of variables, not just polynomials of hypergeometric type,
and it extends to higher dimensions.
The functions in the Askey scheme are just those hypergeometric polynomials that arise as the expansion coef\/f\/icients relating two
separable eigenbases that are {\it both} of hypergeometric type. Thus, there are some
contractions which do not f\/it in the Askey scheme since the physical system fails to have such a~pair of separable eigenbases. In another paper we will analyze the Laplace 2nd order
conformally superintegrable systems, determine which of them is exactly solvable or quasi-exactly solvable,
identify the spaces of polynomials that arise and examine their behavior under contraction.

\subsection*{Acknowledgements}
This work was partially supported by a grant from the Simons Foundation (\# 208754 to Willard Miller~Jr).

\pdfbookmark[1]{References}{ref}
\LastPageEnding


\begin{thebibliography}{99}
\footnotesize\itemsep=0pt

\bibitem{Bocher}
B\^ocher M., Ueber die Reihenentwickelungen der Potentialtheorie, B.G.~Teubner,
 Leipzig, 1894.

\bibitem{Bromwich}
Bromwich T.J.I., Quadratic forms and their classif\/ication by means of
 invariant-factors, Cambridge University Press, Cambridge, 1906.

\bibitem{CapelKress}
Capel J.J., Kress J.M., Invariant classif\/ication of second-order conformally
 f\/lat superintegrable systems, \href{http://dx.doi.org/10.1088/1751-8113/47/49/495202}{\textit{J.~Phys.~A: Math. Theor.}} \textbf{47}
 (2014), 495202, 33~pages, \href{http://arxiv.org/abs/1406.3136}{arXiv:1406.3136}.

\bibitem{CKP2015}
Capel J.J., Kress J.M., Post S., Invariant classif\/ication and limits of
 maximally superintegrable systems in 3{D}, \href{http://dx.doi.org/10.3842/SIGMA.2015.038}{\textit{SIGMA}} \textbf{11} (2015),
 038, 17~pages, \href{http://arxiv.org/abs/1501.06601}{arXiv:1501.06601}.

\bibitem{DASK2007}
Daskaloyannis C., Tanoudis Y., Quantum superintegrable systems with quadratic
 integrals on a two dimensional manifold, \href{http://dx.doi.org/10.1063/1.2746132}{\textit{J.~Math. Phys.}} \textbf{48}
 (2007), 072108, 22~pages, \href{http://arxiv.org/abs/math-ph/0607058}{math-ph/0607058}.

\bibitem{EVAN}
Evans N.W., Super-integrability of the {W}internitz system, \href{http://dx.doi.org/10.1016/0375-9601(90)90611-Q}{\textit{Phys.
 Lett.~A}} \textbf{147} (1990), 483--486.

\bibitem{FORDY}
Fordy A.P., Quantum super-integrable systems as exactly solvable models,
 \href{http://dx.doi.org/10.3842/SIGMA.2007.025}{\textit{SIGMA}} \textbf{3} (2007), 025, 10~pages, \href{http://arxiv.org/abs/math-ph/0702048}{math-ph/0702048}.

\bibitem{HKMS}
Heinonen R., Kalnins E.G., Miller Jr. W., Subag E., Structure relations and
 {D}arboux contractions for 2{D} 2nd order superintegrable systems,
 \href{http://dx.doi.org/10.3842/SIGMA.2015.043}{\textit{SIGMA}} \textbf{11} (2015), 043, 33~pages, \href{http://arxiv.org/abs/1502.00128}{arXiv:1502.00128}.

\bibitem{Wigner}
In\"on\"u E., Wigner E.P., On the contraction of groups and their
 representations, \textit{Proc. Nat. Acad. Sci. USA} \textbf{39} (1953),
 510--524.

\bibitem{Pog01}
Izmest'ev A.A., Pogosyan G.S., Sissakian A.N., Winternitz P., Contractions of
 {L}ie algebras and separation of variables, \href{http://dx.doi.org/10.1088/0305-4470/29/18/024}{\textit{J.~Phys.~A: Math. Gen.}}
 \textbf{29} (1996), 5949--5962.

\bibitem{Pog96}
Izmest'ev A.A., Pogosyan G.S., Sissakian A.N., Winternitz P., Contractions of
 {L}ie algebras and the separation of variables: interbase expansions,
 \href{http://dx.doi.org/10.1088/0305-4470/34/3/314}{\textit{J.~Phys.~A: Math. Gen.}} \textbf{34} (2001), 521--554.

\bibitem{KKM20041}
Kalnins E.G., Kress J.M., Miller Jr. W., Second-order superintegrable systems
 in conformally f\/lat spaces. {I}.~{T}wo-dimensional classical structure
 theory, \href{http://dx.doi.org/10.1063/1.1897183}{\textit{J.~Math. Phys.}} \textbf{46} (2005), 053509, 28~pages.

\bibitem{KKM20041II}
Kalnins E.G., Kress J.M., Miller Jr. W., Second order superintegrable systems
 in conformally f\/lat spaces. {II}.~{T}he classical two-dimensional {S}t\"ackel
 transform, \href{http://dx.doi.org/10.1063/1.1894985}{\textit{J.~Math. Phys.}} \textbf{46} (2005), 053510, 15~pages.

\bibitem{KKM20041III}
Kalnins E.G., Kress J.M., Miller Jr. W., Second order superintegrable systems
 in conformally f\/lat spaces. {III}.~{T}hree-dimensional classical structure
 theory, \href{http://dx.doi.org/10.1063/1.2037567}{\textit{J.~Math. Phys.}} \textbf{46} (2005), 103507, 28~pages.

\bibitem{KKM20041IV}
Kalnins E.G., Kress J.M., Miller Jr. W., Second order superintegrable systems
 in conformally f\/lat spaces. {IV}.~{T}he classical 3{D} {S}t\"ackel transform
 and 3{D} classif\/ication theory, \href{http://dx.doi.org/10.1063/1.2191789}{\textit{J.~Math. Phys.}} \textbf{47} (2006),
 043514, 26~pages.

\bibitem{KKM20041V}
Kalnins E.G., Kress J.M., Miller Jr. W., Second order superintegrable systems
 in conformally f\/lat spaces. V.~Two- and three-dimensional quantum systems,
 \href{http://dx.doi.org/10.1063/1.2337849}{\textit{J.~Math. Phys.}} \textbf{47} (2006), 093501, 25~pages.

\bibitem{KKM20041VI}
Kalnins E.G., Kress J.M., Miller Jr. W., Nondegenerate 2{D} complex {E}uclidean
 superintegrable systems and algebraic varieties, \href{http://dx.doi.org/10.1088/1751-8113/40/13/008}{\textit{J.~Phys.~A: Math.
 Theor.}} \textbf{40} (2007), 3399--3411, \href{http://arxiv.org/abs/0708.3044}{arXiv:0708.3044}.

\bibitem{Laplace2011}
Kalnins E.G., Kress J.M., Miller Jr. W., Post S., Laplace-type equations as
 conformal superintegrable systems, \href{http://dx.doi.org/10.1016/j.aam.2009.11.014}{\textit{Adv. in Appl. Math.}} \textbf{46}
 (2011), 396--416, \href{http://arxiv.org/abs/0908.4316}{arXiv:0908.4316}.

\bibitem{KKMW}
Kalnins E.G., Kress J.M., Miller Jr. W., Winternitz P., Superintegrable systems
 in {D}arboux spaces, \href{http://dx.doi.org/10.1063/1.1619580}{\textit{J.~Math. Phys.}} \textbf{44} (2003), 5811--5848,
 \href{http://arxiv.org/abs/math-ph/0307039}{math-ph/0307039}.

\bibitem{KKMP}
Kalnins E.G., Kress J.M., Pogosyan G.S., Miller Jr. W., Completeness of
 superintegrability in two-dimensional constant-curvature spaces,
 \href{http://dx.doi.org/10.1088/0305-4470/34/22/311}{\textit{J.~Phys.~A: Math. Gen.}} \textbf{34} (2001), 4705--4720,
 \href{http://arxiv.org/abs/math-ph/0102006}{math-ph/0102006}.

\bibitem{KM2014}
Kalnins E.G., Miller Jr. W., Quadratic algebra contractions and second-order
 superintegrable systems, \href{http://dx.doi.org/10.1142/S0219530514500377}{\textit{Anal. Appl. (Singap.)}} \textbf{12} (2014),
 583--612, \href{http://arxiv.org/abs/1401.0830}{arXiv:1401.0830}.

\bibitem{KMP2007a}
Kalnins E.G., Miller Jr. W., Post S., Wilson polynomials and the generic
 superintegrable system on the 2-sphere, \href{http://dx.doi.org/10.1088/1751-8113/40/38/005}{\textit{J.~Phys.~A: Math. Theor.}}
 \textbf{40} (2007), 11525--11538.

\bibitem{KMP2008}
Kalnins E.G., Miller Jr. W., Post S., Models for quadratic algebras associated
 with second order superintegrable systems in 2{D}, \href{http://dx.doi.org/10.3842/SIGMA.2008.008}{\textit{SIGMA}} \textbf{4}
 (2008), 008, 21~pages, \href{http://arxiv.org/abs/0801.2848}{arXiv:0801.2848}.

\bibitem{CCM}
Kalnins E.G., Miller Jr. W., Post S., Coupling constant metamorphosis and
 {$N$}th-order symmetries in classical and quantum mechanics,
 \href{http://dx.doi.org/10.1088/1751-8113/43/3/035202}{\textit{J.~Phys.~A: Math. Theor.}} \textbf{43} (2010), 035202, 20~pages,
 \href{http://arxiv.org/abs/0908.4393}{arXiv:0908.4393}.

\bibitem{KMP2014}
Kalnins E.G., Miller Jr. W., Post S., Contractions of 2{D} 2nd order quantum
 superintegrable systems and the {A}skey scheme for hypergeometric orthogonal
 polynomials, \href{http://dx.doi.org/10.3842/SIGMA.2013.057}{\textit{SIGMA}} \textbf{9} (2013), 057, 28~pages,
 \href{http://arxiv.org/abs/1212.4766}{arXiv:1212.4766}.

\bibitem{Proceedings}
Kalnins E.G., Miller Jr. W., Reid G.J., Separation of variables for complex
 {R}iemannian spaces of constant curvature.~{I}. {O}rthogonal separable
 coordinates for {${\rm S}_{n{\bf C}}$} and {${\rm E}_{n{\bf C}}$},
 \href{http://dx.doi.org/10.1098/rspa.1984.0075}{\textit{Proc. Roy. Soc. London Ser.~A}} \textbf{394} (1984), 183--206.

\bibitem{ArXiv2016}
Kalnins E.G., Miller Jr. W., Subag E., B\^ocher contractions of conformally
 superintegrable {L}aplace equations: detailed computations,
 \href{http://arxiv.org/abs/1601.02876}{arXiv:1601.02876}.

\bibitem{KMS2016}
Kalnins E.G., Miller Jr. W., Subag E., Laplace equations, conformal
 superintegrability and {B}\^ocher contractions, \textit{Acta Polytechnica},
 {t}o appear, \href{http://arxiv.org/abs/1510.09067}{arXiv:1510.09067}.

\bibitem{Koenigs}
Koenigs G., Sur les g\'eod\'esiques a int\'egrales quadratiques, in Darboux~G.,
 Lecons sur la th\'eorie g\'en\'erale des surfaces et les applications
 geom\'etriques du calcul inf\/initesimal, Vol.~4, Chelsea, New York, 1972,
 368--404.

\bibitem{Kress2007}
Kress J.M., Equivalence of superintegrable systems in two dimensions,
 \href{http://dx.doi.org/10.1134/S1063778807030167}{\textit{Phys. Atomic Nuclei}} \textbf{70} (2007), 560--566.

\bibitem{LM2014}
Miller Jr. W., Li Q., Wilson polynomials/functions and intertwining operators
 for the generic quantum superintegrable system on the 2-sphere,
 \href{http://dx.doi.org/10.1088/1742-6596/597/1/012059}{\textit{J.~Phys. Conf. Ser.}} \textbf{597} (2015), 012059, 11~pages,
 \href{http://arxiv.org/abs/1411.2112}{arXiv:1411.2112}.

\bibitem{MPW2013}
Miller Jr. W., Post S., Winternitz P., Classical and quantum superintegrability
 with applications, \href{http://dx.doi.org/10.1088/1751-8113/46/42/423001}{\textit{J.~Phys.~A: Math. Theor.}} \textbf{46} (2013),
 423001, 97~pages, \href{http://arxiv.org/abs/1309.2694}{arXiv:1309.2694}.

\bibitem{NP}
Nesterenko M., Popovych R., Contractions of low-dimensional {L}ie algebras,
 \href{http://dx.doi.org/10.1063/1.2400834}{\textit{J.~Math. Phys.}} \textbf{47} (2006), 123515, 45~pages,
 \href{http://arxiv.org/abs/math-ph/0608018}{math-ph/0608018}.

\bibitem{DLMF}
N{IST} digital library of mathematical functions,
 available at \url{http://dlmf.nist.gov/}.


\bibitem{P2011}
Post S., Models of quadratic algebras generated by superintegrable systems in
 2{D}, \href{http://dx.doi.org/10.3842/SIGMA.2011.036}{\textit{SIGMA}} \textbf{7} (2011), 036, 20~pages, \href{http://arxiv.org/abs/1104.0734}{arXiv:1104.0734}.

\bibitem{TTW2001}
Tempesta P., Turbiner A.V., Winternitz P., Exact solvability of superintegrable
 systems, \href{http://dx.doi.org/10.1063/1.1386927}{\textit{J.~Math. Phys.}} \textbf{42} (2001), 4248--4257,
 \href{http://arxiv.org/abs/hep-th/0011209}{hep-th/0011209}.

\bibitem{SCQS}
Tempesta P., Winternitz P., Harnad J., Miller W., Pogosyan G., Rodriguez M.
 (Editors), Superintegrability in classical and quantum systems, \textit{CRM
 Proceedings and Lecture Notes}, Vol.~37, Amer. Math. Soc., Providence, RI,
 2004.

\bibitem{WW}
Weimar-Woods E., The three-dimensional real {L}ie algebras and their
 contractions, \href{http://dx.doi.org/10.1063/1.529222}{\textit{J.~Math. Phys.}} \textbf{32} (1991), 2028--2033.

\end{thebibliography}
\end{document}